\documentclass[structabstract]{aa}  
\usepackage{natbib,graphicx,amsmath}

\usepackage{txfonts}
\newcommand{\teff}{T$_{\rm eff}$}
\newcommand{\logg}{$\log~g$}

\newcommand{\meta}{${\rm [M/H]}$}

\newcommand{\kms}{km~s$^{-1}$}
\newcommand{\vrad}{${\rm V_{rad}}$}
\newcommand{\myoldgradient}{$\partial \meta / \partial Z=-0.14 \pm 0.05$ dex~kpc$^{-1}$}
\newcommand{\metaDgrad}{$\partial \meta / \partial D$}

\newcommand{\Nsxt}{219}
\newcommand{\Nscl}{166}
\newcommand{\Nfnx}{86}
\newcommand{\Ncar}{208}
\newcommand{\Nforeground}{679}
\begin{document}
%
   

    \title{ Through thick and thin:  Structure of the Galactic thick disc from extragalactic surveys\thanks{Based on observations collected at the European Southern Observatory at Paranal, Chile, ESO Large Programme 171.B-0588 (DART) and 171.B-0520(A).}}

   \author{G.~Kordopatis
          \inst{1}
          \and
          V.~Hill\inst{2}
                    \and
          M.~Irwin\inst{1}
           \and
           G.~Gilmore\inst{1} 
           \and
          R.~F.~G.~Wyse\inst{3}
            \and
             E.~Tolstoy\inst{4}
          \and
          P.~de~Laverny\inst{2}
                    \and				
          A.~Recio-Blanco\inst{2}
	 \and 
	 G.~Battaglia\inst{5}
	 	 \and
	 E.~Starkenburg\inst{6}
          }

   \institute{Institute for Astronomy, University of Cambridge,
              Madingley road, CB3 0HA, Cambridge, UK\\
              \email{gkordo@ast.cam.ac.uk}
         \and
 Laboratoire Lagrange, UMR 7293, Universit\'e de Nice Sophia Antipolis, CNRS, Observatoire de la C\^ote d'Azur, BP 4229, 06304, Nice, France
 	\and
	{Johns Hopkins University, 3400~N Charles Street, Baltimore, MD~21218, USA}
 	\and
	{Kapteyn Astronomical Institute, University of Groningen, PO Box 800, NL-9700 AV Groningen, the Netherlands}
          \and
	{INAF-Osservatorio Astronomico di Bologna, via Ranzani 1, I-40127 Bologna, Italy }
	\and
	{Department of Physics and Astronomy, University of Victoria, PO Box 3055, STN CSC, Victoria BC V8W 3P6, Canada }
	}          

   \date{Received April 12, 2012; accepted }

 
  \abstract
   {We aim to understand the accretion history of the Milky Way by exploring the vertical and radial properties of the Galactic thick disc.}
     {We study the chemical and kinematic properties of roughly a thousand spectra of faint magnitude foreground Galactic stars observed serendipitously during extra-galactic surveys  in four lines-of-sight: three in the southern Galactic hemisphere (surveys of the Carina, Fornax and Sculptor dwarf spheroidal galaxies) and one in  the northern Galactic hemisphere (a survey of the Sextans dwarf spheroidal galaxy). The foreground stars span distances up to  $\sim3$~kpc from the Galactic plane and Galactocentric radii up to 11~kpc. } 
  {The stellar atmospheric parameters (effective temperature, surface gravity, metallicity) are obtained by an automated parameterisation pipeline and the distances of the stars are then derived by a projection of the atmospheric parameters on a set of theoretical isochrones using a Bayesian approach. The metallicity gradients are estimated for each line-of-sight and compared with predictions from the Besan\c{c}on model of the Galaxy, in order to test the chemical structure of the thick disc. Finally, we  use the radial velocities in each line-of-sight to derive a proxy   for either the azimuthal or the vertical component of the orbital velocity of the stars.}  
  {  Only three lines-of-sight have a sufficient number of foreground stars for a robust analysis. Towards Sextans in the Northern Galactic hemisphere and Sculptor in the South, we measure a consistent decrease in mean metallicity with height from the Galactic plane, suggesting a chemically symmetric thick disc. This decrease can either be due to an intrinsic thick disc metallicity gradient, or simply due to a change in the thin disc/thick disc population ratio and no intrinsic metallicity gradients for the thick disc. We favour the latter explanation.  In contrast, we find evidence of an unpredicted metal-poor population in the direction of Carina. This population was earlier detected by Wyse et al. (2006), but our more detailed analysis provides robust estimates of its location  ($|Z|<1$~kpc), metallicity ($-2<$[M/H]$<-1$~dex) and azimuthal  orbital velocity ($V_\phi \sim 120$~\kms).  } 
{Given the chemo-dynamical properties of the over-density towards the Carina line-of-sight, we suggest that it represents the metal-poor tail of the canonical thick disc.
In spite of the small number of stars available, we suggest that this metal-weak thick disc   follows the often suggested canonical thick disc velocity-metallicity correlation of $\partial V_\phi / \partial$[M/H]$\sim 40-50$~\kms~dex$^{-1}$. 
}


   \maketitle
%

\section{Introduction}
Surveys of external  galaxies suggest that thick discs are inherent structures in most disc galaxies \citep{Yoachim08,VanderKruit11}. The existence of such a structure for the Milky Way was proposed thirty years ago \citep{Gilmore83}, but its origin remains uncertain, and many scenarios have been proposed for its formation, either through internal processes or due to external accretion or trigger \citep[][ and references therein]{Abadi03, Brook07, Villalobos08, Loebman11, Rix13}. 

Most of the stellar spectroscopic and photometric surveys have shown that the Galactic thick disc is mainly composed by old stars \citep[older than $\sim 10$~Gyr,][]{Gilmore95, Fuhrmann08}, of intermediate metallicity \citep[\meta $\sim-0.5$~dex,][]{Gilmore85,Bensby07,Kordopatis11b}. In addition, its  stars at the solar cylinder have a ratio of $\alpha$-elements to iron abundances ($[\alpha/\rm{Fe}]$), which is enhanced compared to the $[\alpha/\rm{Fe}]$ abundances of thin disc stars \citep[e.g.:][]{Bensby05,Mishenina06, Reddy06, Fuhrmann08}, which suggests that the thick disc stars were formed on relatively short timescales ($\sim 1$~Gyr), thus offering us the possibility of deciphering the merging history of our Galaxy back to redshifts of $z\sim 1.5-2$.

The amplitudes of vertical and radial gradients in the properties of the 
thick disc place very strong constraints on  Galaxy formation
mechanisms, and there has been significant recent effort towards their measurement. However, the recent results from the SDSS \citep{York00}, SEGUE \citep{Yanny09} and 
RAVE \citep{Steinmetz06} surveys have been obtained either by photometry, from low resolution
spectra, or only for bright, local stars, and are hence limited in either  
accuracy or volume sampled
\citep{Lee11,Chen11,Cheng12,Schlesinger11,Ruchti11}.

Prior to the first results of the Gaia-ESO large observing program \citep{Gilmore12}, the only intermediate resolution spectroscopic survey studying the thick disc far from the solar neighbourhood is that of \citet{Kordopatis11b}. In that study, the authors determined the stellar parameters of roughly 700 stars, from spectra covering the wavelength range of the infrared ionised calcium triplet (IR CaII, $\lambda \sim 8500$~\AA), observed towards the Galactic coordinates $(l,b)\sim (277^\circ,47^\circ)$ and probing distances  from the Galactic plane  of up to 8~kpc.  They found a vertical metallicity gradient of \myoldgradient,  the value of which  could be explained  simply by the changing ratio between thin and thick disc stars as one moves away from the Galactic plane, without any intrinsic vertical  metallicity gradient for the stars of the thick disc. Combining this result with their results for the vertical gradients in Galactocentric rotational velocity and with their finding of no  metallicity dependence of the scale-height and
scale-length of the thick disc,  these authors proposed that the dominant formation mechanisms that could explain their observations were either dynamical heating of a pre-existing thin disc or accretion of a gas-rich satellite.

Here we expand upon the earlier study of \citet{Kordopatis11b} with four additional lines-of-sight (los), analysing the foreground Galactic stars of the extragalactic stellar survey of the {\it Dwarf Abundances and Radial velocities Team} \citep[DART,][]{Tolstoy06} towards the Sculptor, Fornax and Sextans dwarf spheroidal galaxies (dSph), together with  the foreground stars observed as part of the ESO large program 171.B-0520(A) to study the Carina dSph.   This paper is structured as follows: In Sect.~\ref{Sect:parameterisation} we present  the dataset and explain how the stellar atmospheric parameters were  obtained using an automated pipeline. In Sect.~\ref{Sect:Distances_and_comparisons} we compute spectroscopic distances for all the stars (both members of the dSph and 
in the foreground), describe how we select our final sample of foreground Galactic stars, and derive an estimator for the orbital azimuthal
velocity of the stars. Then, in Sect.~\ref{Sect:Results}, we compare these results 
with predictions from the Besan\c{c}on model in  each
line-of-sight to aid the identification of substructure.  Finally,  in Sect.~\ref{Sect:Discussion}  we discuss possible origins for the over-density we find in the line-of-sight towards Carina, and Sect.~\ref{Sect:conclusions} presents our conclusions.

\begin{table*}[htdp]
\caption{Lines-of-sight and available spectra}
\label{Tab:Observations}
\begin{center}
\begin{tabular}{cccccccc}
\hline \hline


line-of-sight & dwarf  galaxy & $<A_V>$ & $<A_I>$ &  \vrad~ dSph  & N spectra &  N objects  \\
	&			& (mag)		& (mag)		&	(\kms)	& &  \\ \hline

I:~($l=287.5^\circ, b=-83.2^\circ$) & Sculptor & 0.061 &0.036& $110 \pm 10.1$ & 2393 & 1488 \\
II:~($l=243.5^\circ, b=42.3^\circ$) & Sextans & 0.164   &0.096& $226 \pm 8.4$ & 1512 & 1022 \\
III:~($l=237.1^\circ, b=-65.7^\circ$) & Fornax & 0.074   &0.043& $54.1 \pm 11.4$ & 1352 & 1060 \\
IV:~($l=260.1^\circ, b=-22.2^\circ$) & Carina & 0.211   &0.123& $223.9 \pm 7.5$ & 1319 & 939 \\
\hline \hline

\end{tabular}
\end{center}
\label{Tab:resume_los}
\end{table*}%

\section{Measurement of the stellar atmospheric parameters}
\label{Sect:parameterisation}

\subsection{Description of the datasets}
\label{Subsect:data}
In this work, we made extensive use of the spectroscopic data from the extragalactic survey DART and from the observing program 171.B-0520(A). 

The large ESO observing program DART aimed to determine the chemical and dynamical properties of three dwarf spheroidal galaxy (dSph) satellites of the Milky Way.  This survey  obtained medium- ($R\sim6500$) and high-  ($R\sim20~000$) resolution spectra in lines-of-sight towards Sculptor ($l \sim 287.5^\circ, b\sim -83.2^\circ$), Sextans ($l=243.5^\circ, b=42.3^\circ$) and Fornax ($l=237.1^\circ, b=-65.7^\circ$). In this paper we used the subset of medium resolution spectra, taken around the IR CaII triplet at $\sim 8500$~\AA, with the VLT-ESO FLAMES-Giraffe spectrograph at the LR8 setup.  The data-reduction procedure and the techniques for derivation of the radial velocities that we adopted are described in detail in \citet{Battaglia08b}.  The targets observed towards the Carina line-of-sight were observed as part of a second ESO Large Observing Program \citep[see][for a discussion of the Carina member stars]{Koch06} and were  observed with the same spectrograph, and reduced following the same procedure as in \citet{Battaglia08b}.

\begin{figure}[t]
   \centering
\includegraphics[width=0.5\textwidth,,angle=180]{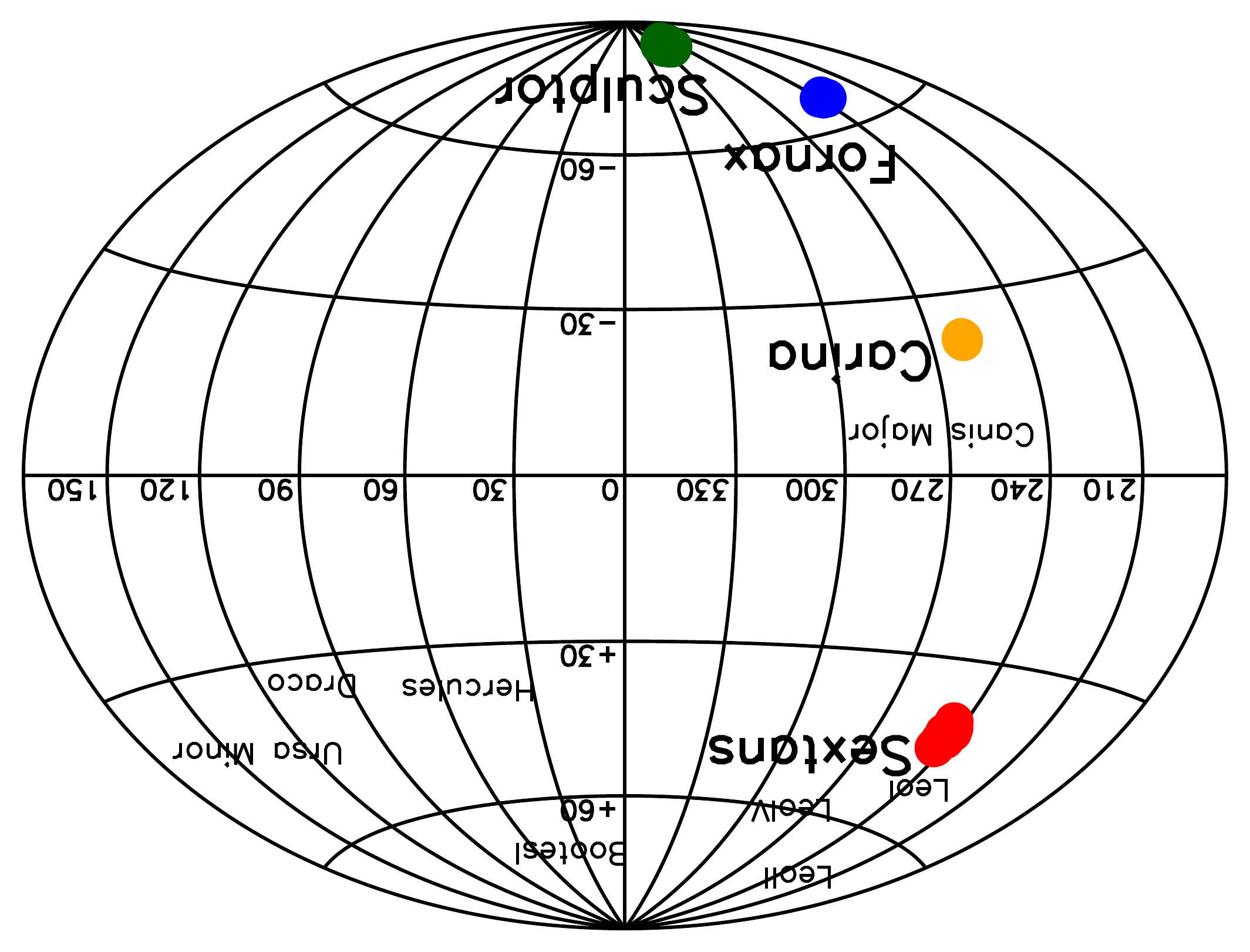}
   \caption{Aitoff projection of the Galactic coordinates of the targets. The lines-of-sight of Sculptor, Sextans, Fornax and Carina are represented in green, red, blue and orange colours, respectively. The positions of other dwarf galaxies, including  Ultra Faint Dwarfs, have also been indicated.}%
      \label{Fig:Coordinates}
    \end{figure}

The total number of targets observed towards the lines-of-sight to Sculptor, Sextans, Fornax and Carina were, respectively, 1488, 1022, 1060 and 939,  for which
repeat observations were gathered for several of them. The entire 
database for these four line-of-sight contains 6576 spectra, consisting of  2393,
1512, 1352 and 1319 spectra, respectively, for each (see
Fig.~\ref{Fig:Coordinates} and Table~\ref{Tab:resume_los}). As will
be shown in the following sections, out of these 6576 spectra, only \Nforeground\ targets represent 
foreground stars (see Sect.~\ref{sect:foregound_selection}).

\subsection{Adopted pipeline for the automated parameterisation}
\label{sect:pipeline}

An updated version of the pipeline presented in \citet{Kordopatis11a} was developed to obtain estimates of the values of the atmospheric parameters of the stars. This procedure consists of using two different algorithms simultaneously, MATISSE \citep{Recio-Blanco06} and DEGAS \citep{Bijaoui12}, to renormalise (iteratively) the spectra, compute the mean signal-to-noise ratio (SNR) per pixel and derive the atmospheric parameters of the observed stars, as well as their associated internal errors. The learning phase for the algorithms is based on a grid of synthetic spectra covering \teff\ from 3000~K to 8000~K, \logg\ from 0 to 5 (cm~s$^{-2}$) and [M/H] from $-5$~dex to $+1.0$~dex. A coupling between the overall metallicity and the $\alpha-$element abundances\footnote{The chemical species considered as $\alpha$-elements are \ion{O}{}, \ion{Ne}{}, \ion{Mg}{}, \ion{Si}{}, \ion{S}{}, \ion{Ar}{}, \ion{Ca}{} and \ion{Ti}{}.}  is assumed according to the commonly observed enhancements  in local metal-poor thick disc stars:
\begin{itemize}
\renewcommand{\labelitemi}{$\bullet$}
\item $[\alpha$/Fe]=0.0 dex for  [M/H] $\geq$0.0 dex
\item $[\alpha$/Fe]=$-0.4 \times $[M/H] dex for $-1\leq$[M/H]$\leq 0$~dex
\item $[\alpha$/Fe]=+0.4 dex for [M/H] $\leq$ $-$1.0  dex.
\end{itemize}

Compared to the previous learning grid used in \cite{Kordopatis11a}, new model spectra have been added, to maintain a constant metallicity step of 0.25~dex across all the parameter space and to minimise the border effects and discretisation problems encountered by 
\citet{Kordopatis11a}. 
These additional spectra have been computed by linearly interpolating the flux between the available nominal spectra from the previous version of the reference grid. After this procedure, the updated pipeline has a nominal grid of 4571 synthetic spectra. 

In addition, the updated pipeline includes the capability of imposing  soft priors, based on the observational strategy and the known selection function of the survey. For example, one can remove from the solution space all the templates which do not correspond to the photometric temperatures expected, based on the colour-magnitude cuts used to select the sample. Removing such templates greatly  reduces the degeneracy problems from which all automated parameterisation algorithms may suffer and hence improve the accuracy of
the results.

The selection functions of the DART and Carina surveys were defined to identify  stars on the Red Giant Branches of the targeted dSph.  The samples observed therefore consist of faint stars ($16<V<20$~mag) with a colour range of $0.6 <(V-I)< 2.1$~mag.  Given these  criteria, and using the photometric temperature calibration relations of \citet{Ramirez05} and \citet{Casagrande10}, the effective temperatures of the stars should be in the range $3500 \leq $\teff $\leq 6500$~K, no matter their metallicity or surface gravity. We therefore ran the pipeline with a prior such that only solutions within this range were retained.

   \begin{figure}[t]
   \centering
\includegraphics[width=0.5\textwidth]{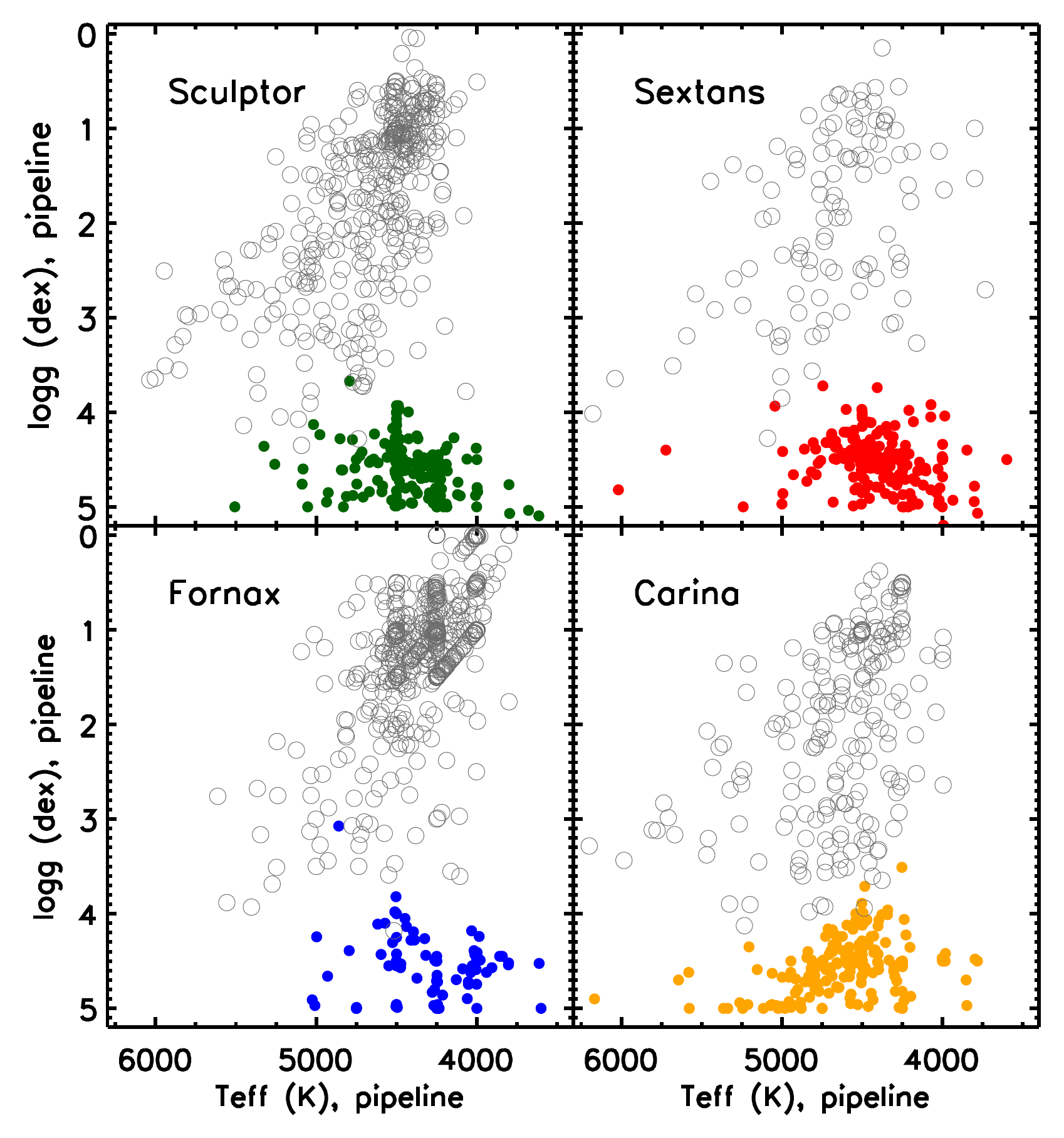}
   \caption{Effective temperature (\teff) -- surface gravity (\logg) diagram representing the parameters measured by the pipeline. The foreground stars  corresponding to the line-of-sight of Sculptor are represented in green, the ones towards Sextans in red, towards Fornax in blue and towards Carina in orange. The background extragalactic stars, as selected in Sect.~\ref{sect:foregound_selection}, are represented in grey open circles.  }%
   \label{Fig:H-R_diagram_pipeline}
    \end{figure}

\onltab{2}{
\begin{table*}
\caption{Atmospheric parameters of the selected foreground stars}\label{tab:atm_parameters}
\begin{tabular}{lcccccccccc}
\hline \hline
ID  & \teff$_p$ & \teff$_i$ &$\Delta$ \teff & \logg$_p$ & \logg$_i$ & $\Delta$ \logg & [M/H]$_p$ & [M/H]$_c$ & $\Delta$ [M/H] \\ 
      & (K) & (K) & (K) & (cm~s$^{-2}$) & (cm~s$^{-2}$) & (cm~s$^{-2}$) & (dex) & (dex) &(dex)  \\ \hline
       \multicolumn{5}{l} {line-of-sight: Sculptor        }  & \multicolumn{5}{r} {Foreground stars=  166                   } \\ \hline
       I-1    &    3998    &    4790    &     119    & 4.80    & 4.64    & 0.23    &-0.72    &-0.44    & 0.14   \\
       I-2    &    3800    &    4579    &     119    & 4.76    & 4.66    & 0.23    &-0.25    &-0.11    & 0.14   \\
       ... \\
 \hline \multicolumn{5}{l} {line-of-sight: Sextans        }  & \multicolumn{5}{r} {Foreground stars=  219                  } \\ \hline 
      II-1    &    4500    &    4535    &      64    & 4.50    & 4.72    & 0.12    &-0.50    &-0.27    & 0.08   \\
      II-2    &    4675    &    4848    &      64    & 4.49    & 4.55    & 0.12    &-0.12    &-0.02    & 0.08   \\
      ...\\
        \hline \multicolumn{5}{l} {line-of-sight: Fornax        }  & \multicolumn{5}{r} {Foreground stars=   86                  } \\ \hline 
     III-1    &    4548    &    4628    &     119    & 4.55    & 4.69    & 0.23    &-0.71    &-0.43    & 0.14   \\
     III-2    &    4447    &    4215    &      64    & 4.05    & 4.75    & 0.12    &-0.81    &-0.49    & 0.08   \\
     ...\\
       \hline \multicolumn{5}{l} {line-of-sight: Carina        }  & \multicolumn{5}{r} {Foreground stars=  208                  } \\ \hline 
      IV-1    &    4793    &    4576    &     119    & 4.58    & 4.69    & 0.23    &-0.75    &-0.46    & 0.14   \\
      IV-2    &    4433    &    4555    &     158    & 4.91    & 4.78    & 0.26    &-1.25    &-0.94    & 0.18   \\
... \\ \hline
\end{tabular}
\tablefoot{The $p$, $i$ and $c$  subscripts refer to the atmospheric parameters coming from the pipeline, projected on the isochrones, and calibrated according to Eq.~\ref{eq:calib_meta}, respectively.  The errors refer to the internal errors. Full table available  at the CDS.}
\end{table*}
}

The output values of the stellar atmospheric parameters of the foreground stars, as selected in Sect.~\ref{sect:foregound_selection}, are listed in the electronic Table~\ref{tab:atm_parameters}.  They are also plotted in the \teff--\logg~diagram of Fig.~\ref{Fig:H-R_diagram_pipeline} together with the parameters of the dSph members for which we had spectra with SNR$>$10~pixel$^{-1}$. The distribution of the points in  this diagram is noisy, especially along the gravity axis, due to the low SNR of the spectra (ranging from $\sim 3$ to 70~pixel$^{-1}$), combined with the poor gravity-sensitivity of the spectral features available in the IR-Ca~II triplet region.  Indeed, as shown in \cite{Kordopatis11a}, surface gravity is poorly constrained for main-sequence dwarfs, due to the lack of sensitivity of the spectral lines to small variations of \logg.  Nevertheless, we 
verified that differentiation between dwarfs and giants can be achieved fairly 
easily, based (among others) on the strength of the \ion{Mg}{I} line at $8806.8$~\AA,  which is more prominent in the spectra of dwarfs \citep[see also][]{Battaglia12}.

Finally, we note that a work in progress comparing metallicities obtained from high-resolution spectra with the results obtained from the present pipeline applied to spectra from the  RAVE survey, of  the same wavelength range and  spectral resolution as the present work \citep[in prep.]{Kordopatis_RAVE}, shows that the pipeline metallicities have to be corrected by a low order polynomial which is a function of \logg\ and \meta. The relation that we used to correct the pipeline metallicity values is the following:
\begin{small}
\begin{equation}
\begin{split}
[M/H]_{\rm p} - [M/H]_{\rm c}=-0.076 - 0.006 * \log g + 0.003*\log^2g \\
- 0.021*[M/H]_{\rm p}*\log g+0.582*[M/H]_{\rm p}+0.205*[M/H]_{\rm p}^2
\end{split}
\label{eq:calib_meta}
\end{equation}
\end{small}
where $[M/H]_{\rm c}$ and $[M/H]_{\rm p}$ are the corrected and pipeline metallicity values.  This calibration suggests a correction of less than 0.1~dex for solar-metallicity dwarfs and $\sim0.35$~dex for giants of metallicity between $-2.5<\meta<0.1$~dex. There is no indication of the need for corrections to the pipeline output for \teff\ or \logg. Both the pipeline metallicities and the calibrated metallicities are listed in the electronic Table~\ref{tab:atm_parameters}, but for  the remainder of this paper, 'metallicity' will  refer to only the calibrated value.\\

  \begin{figure}[t]
   \centering
\includegraphics[width=0.35\textwidth]{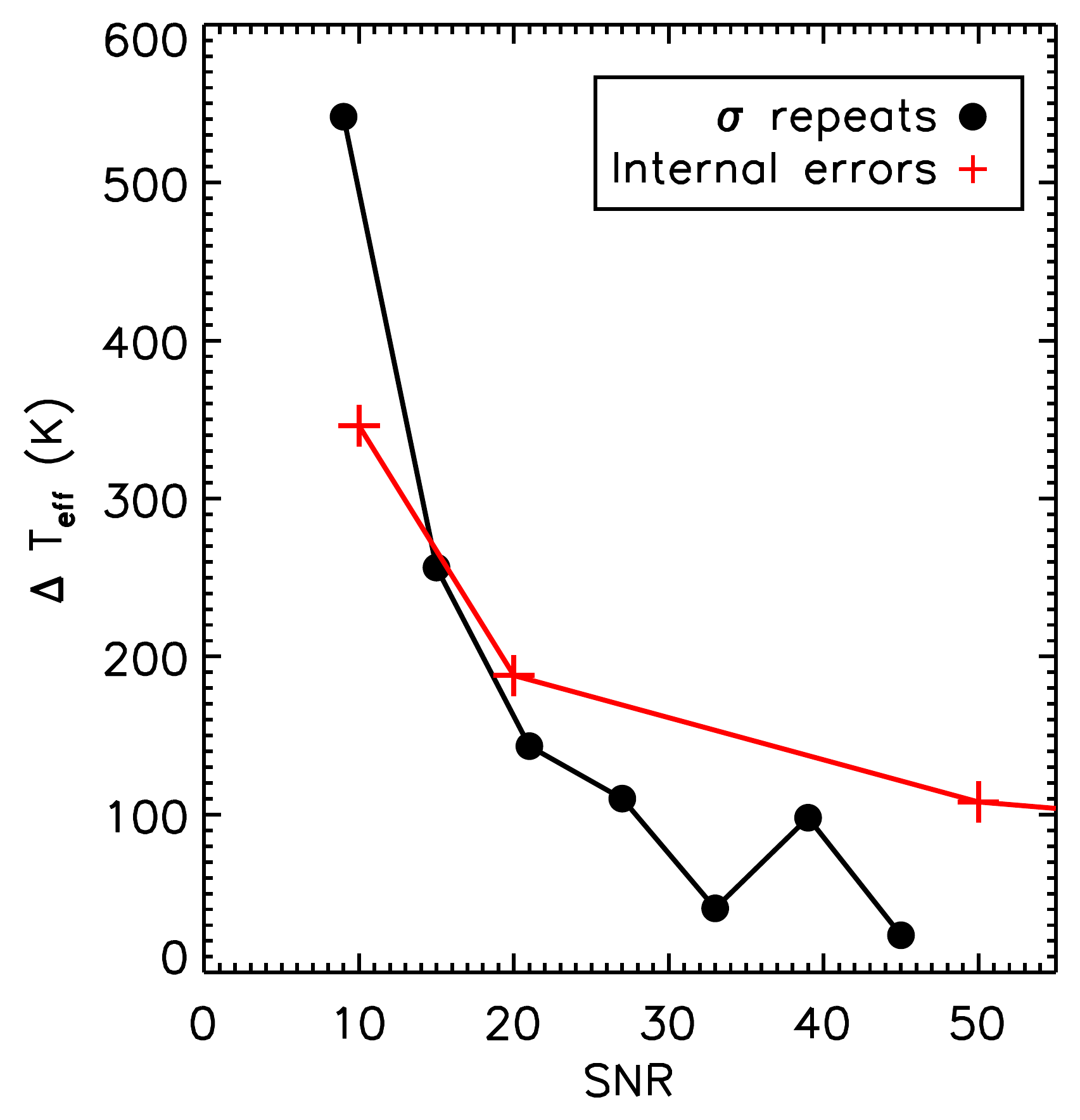}\\
\includegraphics[width=0.35\textwidth]{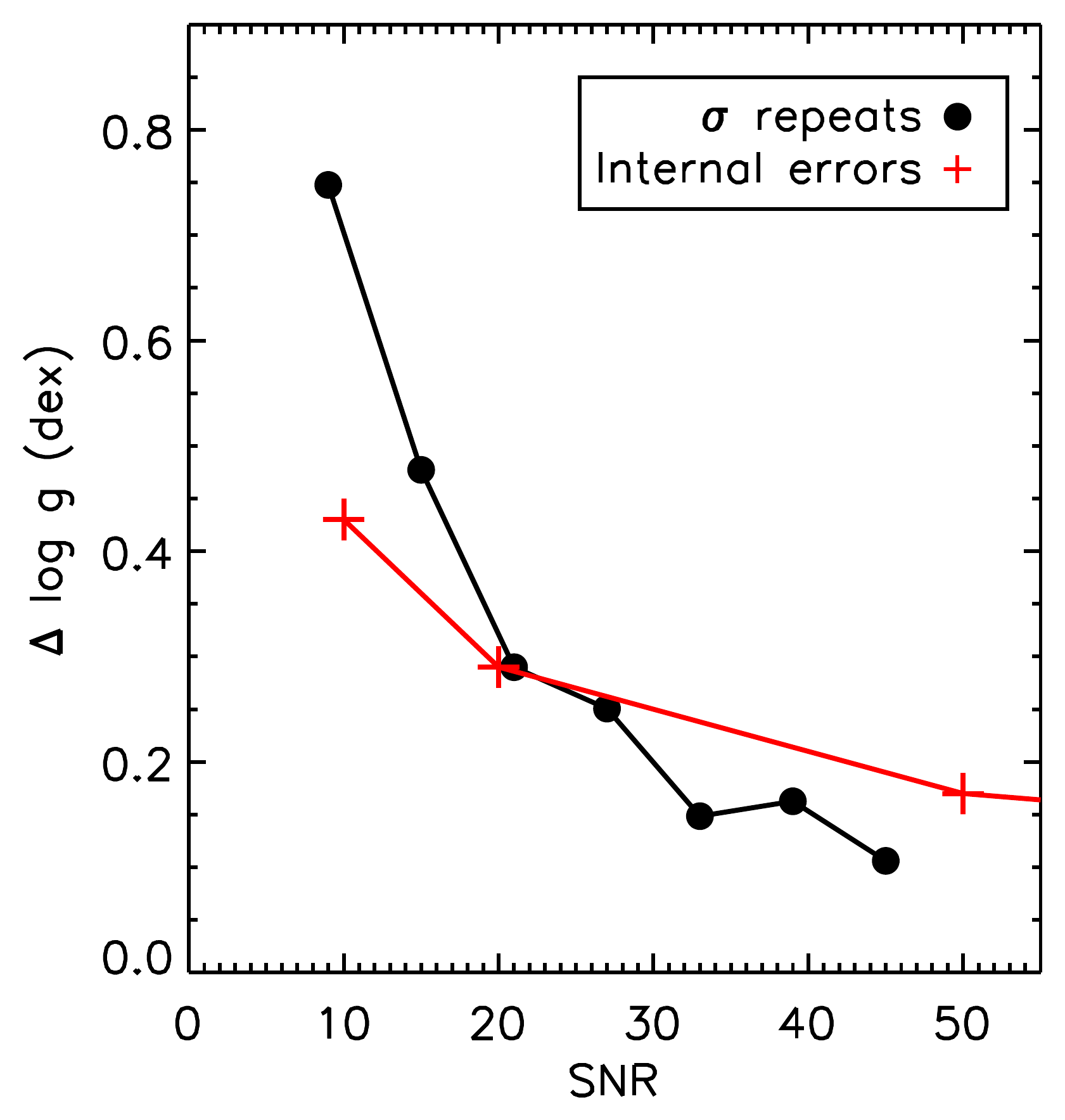}\\
\includegraphics[width=0.35\textwidth]{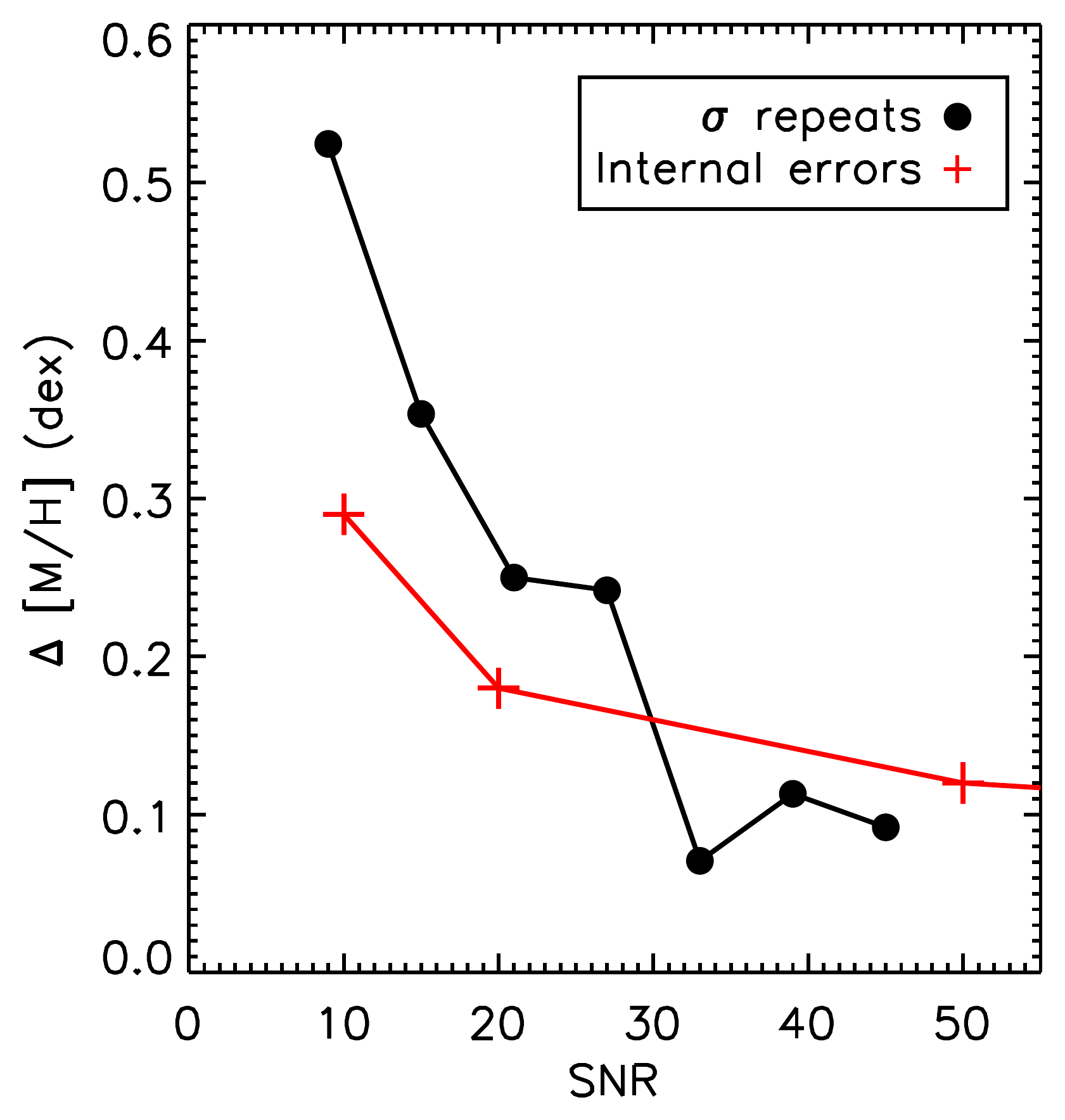}
   \caption{Comparison of the internal uncertainties (in red) with the 70th percentile of the dispersion distribution of the repeated observations of dwarf stars more metal rich than [M/H]$\sim -1.25$~dex for \teff~(top), \logg~(middle) and \meta~(bottom) }%
   \label{Fig:internal_errors}
    \end{figure}

\subsection{Total uncertainties in the estimated parameters}
Based on the repeated observations, we  investigated to which extent the internal uncertainties, purely associated to our pipeline's parameter estimations, were realistic. The internal uncertainties  in a given set of atmospheric parameters derived by the pipeline from a spectrum of a given SNR are estimated from the values given in Table~4 of \cite{Kordopatis11a}. That analysis was based on the results derived from synthetic spectra with white Gaussian noise, and it assumed there were no correlations between different parameters, no spectral degeneracies and Gaussian uncertainties. The validity of these assumptions may be assessed by comparing those uncertainties with the dispersions in parameter values derived from repeated observations of the same star.

The current analysis focuses on foreground dwarf stars and we therefore investigated the repeat observations of those stars that the pipeline identifies as dwarfs. We first measured  the dispersion in the parameter estimations for each star observed multiple times, and associated to that dispersion the mean SNR of the stellar spectra under consideration. Then, for different SNR bins, the 70th~percentile of the dispersion distribution was measured and compared to the internal uncertainties for FGK dwarfs of metallicities down to $-1$~dex at  that SNR.  Figure~\ref{Fig:internal_errors} shows the comparison between the internal uncertainties (red plus signs) and the dispersions in the values of the parameters for the repeats that have been designated as dwarfs, with metallicities higher than $-1.25$~dex (black curves). It is apparent that we slightly over-estimate the errors for the high SNR regimes, and under-estimate them by a factor of less than two at a SNR$\sim 10$~pixel$^{-1}$. The different behaviours of the uncertainties from the repeats and the estimated internal errors seen in the Fig.~\ref{Fig:internal_errors}  are  due in part to the fact that the abscissa of the black curves is not truly the SNR but rather the mean SNR of the spectra under consideration. Additional factors behind the disagreement are that the error ellipsoid has not been taken into consideration when computing the uncertainties, as well as the fact that we are now dealing with observed spectra rather than synthetic ones. Nevertheless, despite these approximations, the agreement is satisfactory.

\begin{table}[tdp]
\caption{Estimation of the external uncertainties. The boundary between dwarfs and giants is at \logg$=3.5$~dex, between hot and cool stars at \teff$=6000$~K and between metal-poor and metal-rich stars at \meta$=-0.5$~dex. \citep[][in prep.]{Kordopatis_RAVE}}
\begin{center}
\begin{tabular}{c|cccc}
\multicolumn{5}{c}{Dwarfs} \\ \hline
Parameter range       &             N   &    $\sigma$(\teff)     & $\sigma$(\logg) &$\sigma$(\meta) \\ \hline
Hot, metal-poor   &      28   &   314 &    0.466 &     0.269 \\
Hot, metal-rich    &  104    &  173   &  0.276 &     0.119  \\
Cool, metal-poor     &     97   &   253   &  0.470 &     0.197  \\
Cool, metal--rich   &   138    &  145 &     0.384 &     0.111 \\ \hline

\multicolumn{5}{c}{ } \\ 
\multicolumn{5}{c}{Giants} \\ \hline
Parameter range         &             N   &    $\sigma$(\teff)     & $\sigma$(\logg) &$\sigma$(\meta) \\ \hline
Hot           & 8     & 263 &    0.423 &     0.300 \\
Cool, metal-poor &     273 &     191 &     0.725 &     0.217 \\
Cool, metal-rich   &   136   &   89 &     0.605 &    0.144 \\ \hline
\end{tabular}
\end{center}
\label{tab:external_errors}
\end{table}%

In order to treat the uncertainties of the pipeline realistically, external errors must be evaluated and added quadratically to the internal ones. For that reason, we have used the external uncertainty estimations used in \citet[in prep.]{Kordopatis_RAVE}, obtained for  a set of roughly 800 stars with SNR$>$50~pixel$^{-1}$, observed by RAVE  at the same wavelength range and spectral resolution, and for which atmospheric parameters were available from  high-resolution spectroscopy. The adopted external uncertainties are given in Table~\ref{tab:external_errors}.

Finally, each star should have only one estimated value of the atmospheric parameters in the database, to avoid biases in the sample distributions. Thus for  the spectra corresponding to the same object, the  mean of the parameter values was computed, weighted by the SNR of the respective spectrum. Hence, for a given parameter value $\theta_{i,s}$, associated with spectrum $s$ and an object $i$, the adopted final parameter value is :

\begin{equation}
\theta_i = \frac {\sum {\rm SNR}_{i,s} \cdot \theta_{i,s}} {\sum SNR_{i,s}}
\end{equation}
The errors $\sigma_{\theta_i}$ of the parameters have been computed as the weighted standard deviation: 
\begin{equation}
\sigma^2_{\theta_i} = \frac {\sum {\rm SNR}^2_{i,s} \cdot \sigma_{\theta_{i,s}}^2} {\sum SNR^2_{i,s}}.
\end{equation}

   \begin{figure}[t]
   \centering
\includegraphics[width=0.48\textwidth]{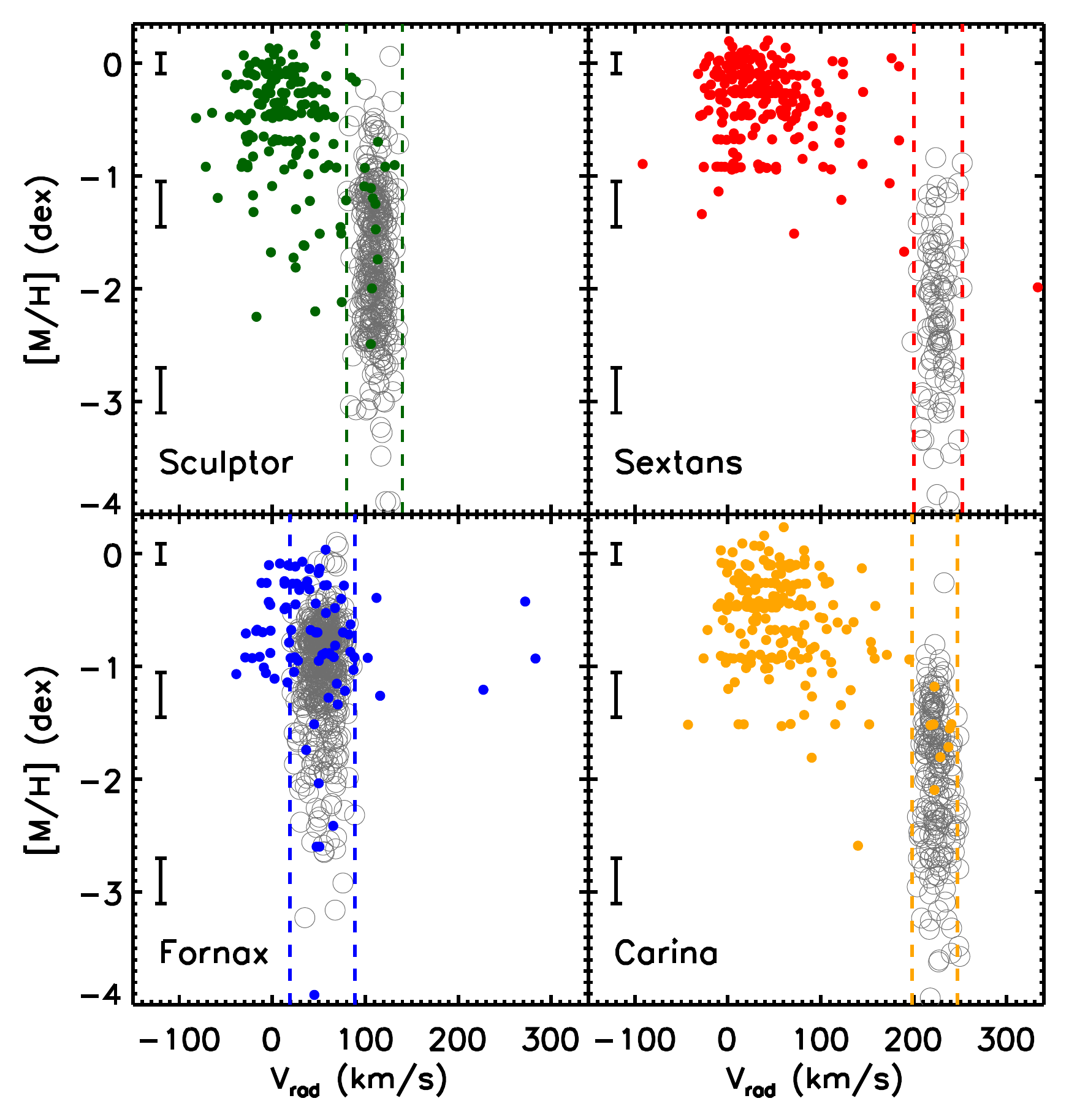}
   \caption{Radial velocity versus overall metallicity for all the lines-of-sight. The colour coding is the same as in Fig.~\ref{Fig:H-R_diagram_pipeline}. Typical internal uncertainties, at different metallicity regimes, and for the whole dataset (including the faint extra-galactic stars plotted as grey empty circles) are represented in the left part of the plots. The dotted vertical lines represent for each line-of-sight the adopted radial velocity range of the stars belonging to each dwarf spheroidal galaxy. }%
   \label{Fig:Vrad_vs_meta}
    \end{figure}

Figure~\ref{Fig:Vrad_vs_meta} illustrates the relation between the radial velocity, \vrad, and the metallicity of the foreground (filled circles) and dSph (open circles) stars in our sample, as they are defined in Sec.~\ref{sect:foregound_selection}. The stars belonging to the dSph are concentrated within narrow radial velocity distributions, and they span low metallicity values, as expected. The mild over-densities at discrete metallicity values that can be noticed for the foreground stars are a signature of the DEGAS algorithm, which tends to find parameter values close to those of the learning grid points.  We emphasise that this discretisation does not affect in any way the accuracy of the parameter estimations, as tests have shown that this  pseudo-discretisation minimises the internal errors of the pipeline \citep{Kordopatis11a}.

\section{Computation of the distances and final sample selection}
\label{Sect:Distances_and_comparisons}

Any investigation of the chemical structure of the thin and thick discs must  first determine stellar distances, and then gradients and distribution functions. In the absence of parallaxes, the best way to estimate distances is by the projection of the stellar atmospheric parameters onto a set of theoretical isochrones to obtain the absolute magnitude. We explain below how we implemented such a procedure to obtain the distances of the targets, then discuss which criteria were used to disentangle the foreground stars from from those belonging to the background dSphs, and
describe how we obtained a proxy for  the azimuthal velocity of the stars.

\subsection{Computation of the distances}
\label{Sect:distances}
   
We used the same procedure as presented in \cite{Kordopatis11b} to obtain, for each star, the line-of-sight distance ($D$), as well as the  Galactocentric radius  in cylindrical coordinates ($R$) and the distance above the Galactic plane ($Z$).

This procedure consists of assigning a weight to each point of a  set of theoretical isochrones, taking into account the derived stellar atmospheric parameters, their total uncertainties (internal and external) and the Bayesian probability of a star to be in a particular evolutionary phase.  It is worth mentioning that in our implementation this probability is  simply  related to the amount of time spent by a star in each evolutionary phase of an  isochrone, and is not related to any priors based on the Galactic stellar population such as was the case in, for example, \cite{Burnett10}.  The most likely absolute magnitude of the star is then obtained by computing the weighted mean of all the absolute magnitudes (in a given band) of the isochrones.

First, we de-reddened the magnitudes of our stars, adopting the extinction values of \citet{Schlegel98}, given in Table~\ref{Tab:resume_los}.  We adopted the same extinction for all the stars in each line-of-sight. This may be justified by the fact that most of the inter-stellar matter is confined close to the Galactic plane, within a scale-height of $\sim 150$~pc \citep{Misiriotis06}, and that most of the observed stars lie beyond a vertical distance of $Z=300$~pc.  
We then generated our own set of isochrones by using the $YYmix2$ interpolation code, based on the Yonsei-Yale ($Y^2$) models \citep[version 2,][]{Demarque04}, combined with the \cite{Lejeune98} colour table, as in \cite{Kordopatis11b}.  We set the youngest age of the isochrones at 2 Gyr, based on the expectation that the observed stars will be  from the old thin disc, the thick disc and the stellar halo.  The interpolated isochrones have a constant age step of 1~Gyr, up to 14~Gyr. In addition, we used the full metallicity range of the $Y^2$ models, ranging from [Fe/H]$=-3.0$~dex to [Fe/H] $= +0.8$~dex. The step in metallicity is constant,
equal to 0.1 dex, smaller than the typical error on the derived metallicities of the observed stars (typically $\sim 0.15-0.2$~dex, see Fig.~\ref{Fig:internal_errors}). The adopted values for the $\alpha-$enhancements at different metallicities are the same as those used for the grid of synthetic spectra described in Sect.~\ref{sect:pipeline}. In the end, a set of 494 isochrones were generated.

  \begin{figure}[t]
   \centering
\includegraphics[width=0.35\textwidth]{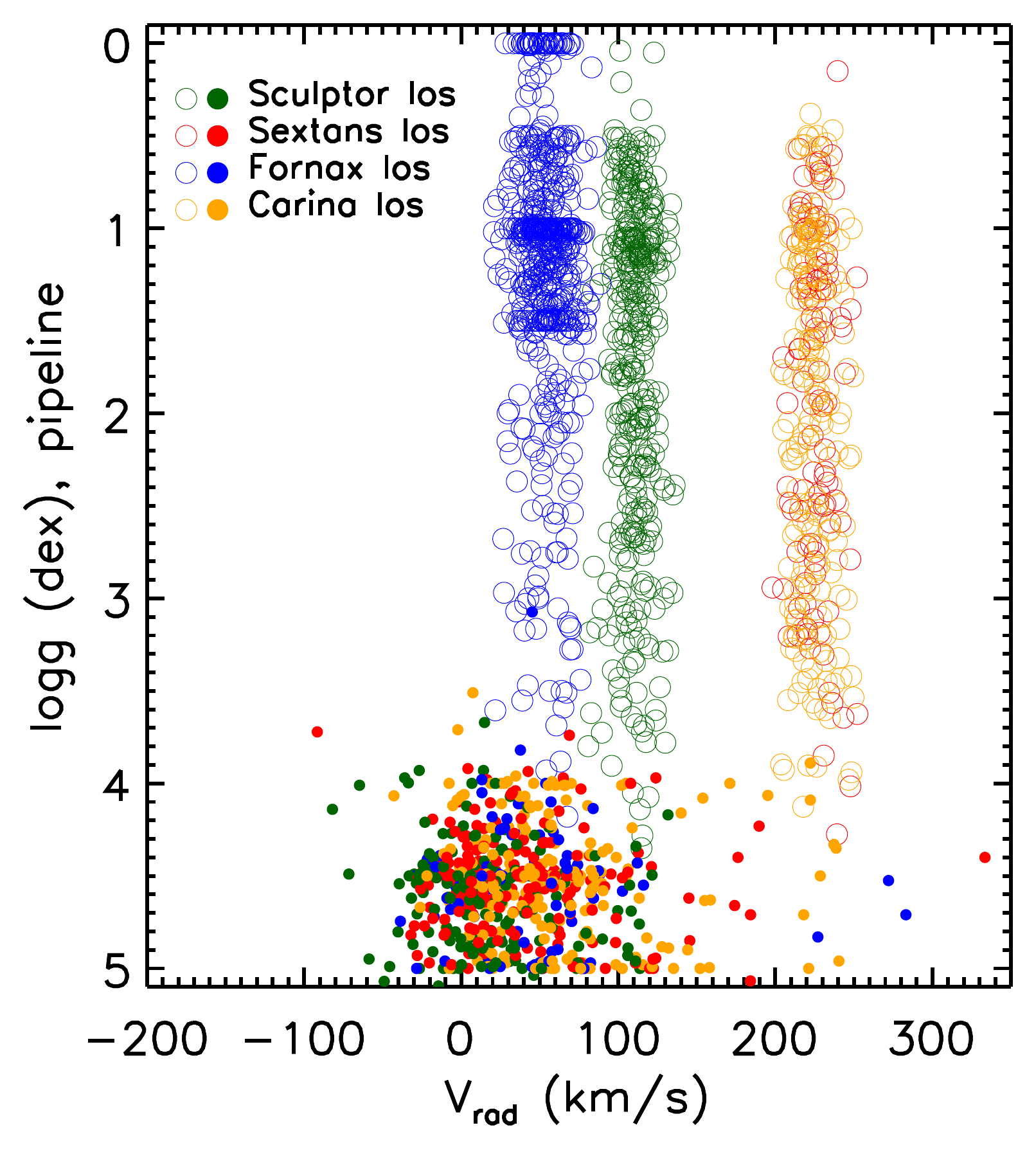}\\
\includegraphics[width=0.35\textwidth]{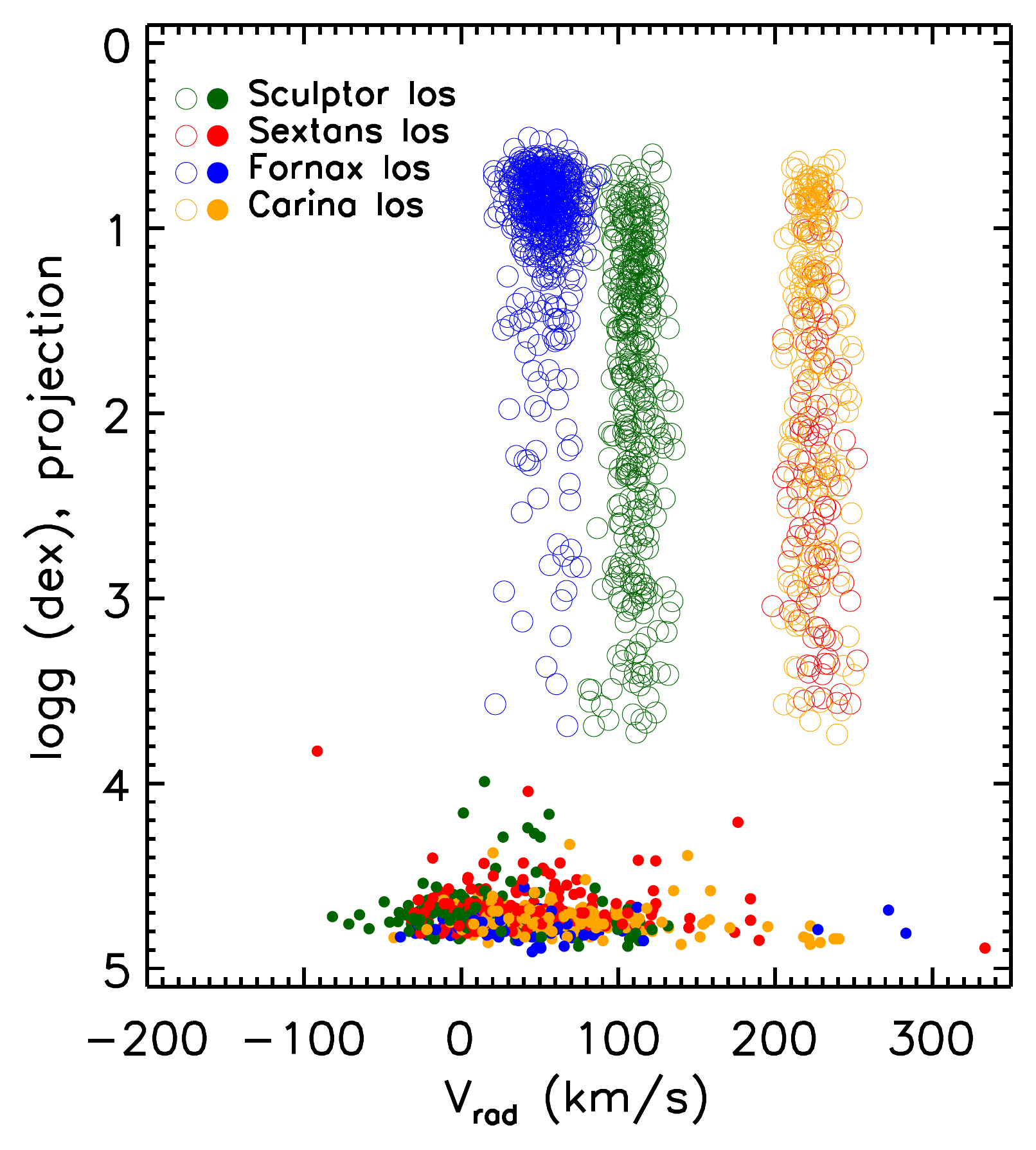}
   \caption{ Radial velocity versus surface gravity, using the results from the parametrisation pipeline (upper panel) and those projected on the Yonsei-Yale isochrones (lower panel). Open symbols correspond to the targets that have the same velocity as the dSph in their line-of-sight and have a projected (log) surface gravity less than 3.75~cm~s$^{-2}$. }%
   \label{Fig:logg_vrad}
    \end{figure}

   \begin{figure}[t]
   \centering
\includegraphics[width=0.5\textwidth]{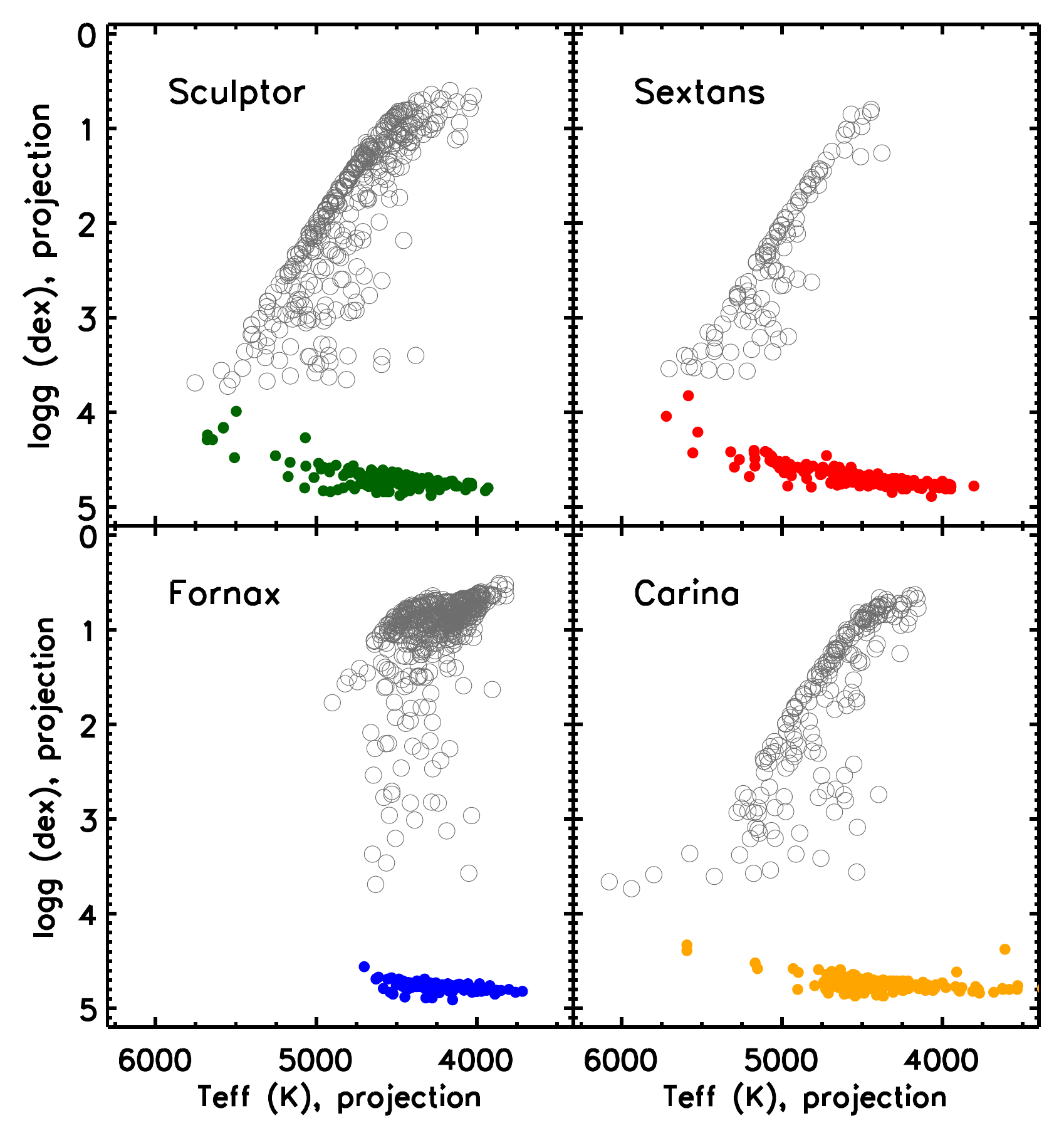}
   \caption{Effective temperature (\teff) -- surface gravity (\logg) diagram using the parameter values obtained after  projection onto isochrones   (see Sect.~\ref{Sect:distances}), for all the observed targets of the DART and Carina surveys.  For each line-of-sight, the filled symbols correspond to the foreground stars that are considered in this study, while the open grey circles are the dSph members. }%
   \label{Fig:H-R_diagram}
    \end{figure}

 Figure~\ref{Fig:logg_vrad} demonstrates the change in the estimates of surface gravities before and after the projection onto the isochrones, as a function of the radial velocity.
 Foreground stars, as selected in Sect.~\ref{sect:foregound_selection}, are plotted in filled circles, and extra-galactic stars as open circles.  It is apparent that projection onto the isochrones makes the distinction between dwarf stars (\logg$> 3.75$~cm~s$^{-2}$) and giants clearer; in the case of stars having a radial velocity equal to that of the dSph, this distinction corresponds to differentiating extra-galactic (giant) stars from foreground (dwarf) ones.  The improvement in the estimation of the atmospheric parameters can also be seen  in Fig.~\ref{Fig:H-R_diagram}, showing the \teff--\logg~diagram of the projected parameters.  Compared to the distribution in Fig.~\ref{Fig:H-R_diagram_pipeline}, the stars for which the pipeline obtained unrealistic parameter combinations (e.g.~\teff$\sim 4200~$K, \logg$\sim 4.0$~cm~s$^{-2}$ and high [M/H]), have been moved by the projection into allowed locations, either the red giant branch or the main sequence.  The values of the projected effective temperatures and surface gravities are listed in the electronic Table~\ref{tab:atm_parameters}, next to the raw pipeline results.

   \begin{figure}[t]
   \centering
\includegraphics[width=0.48\textwidth]{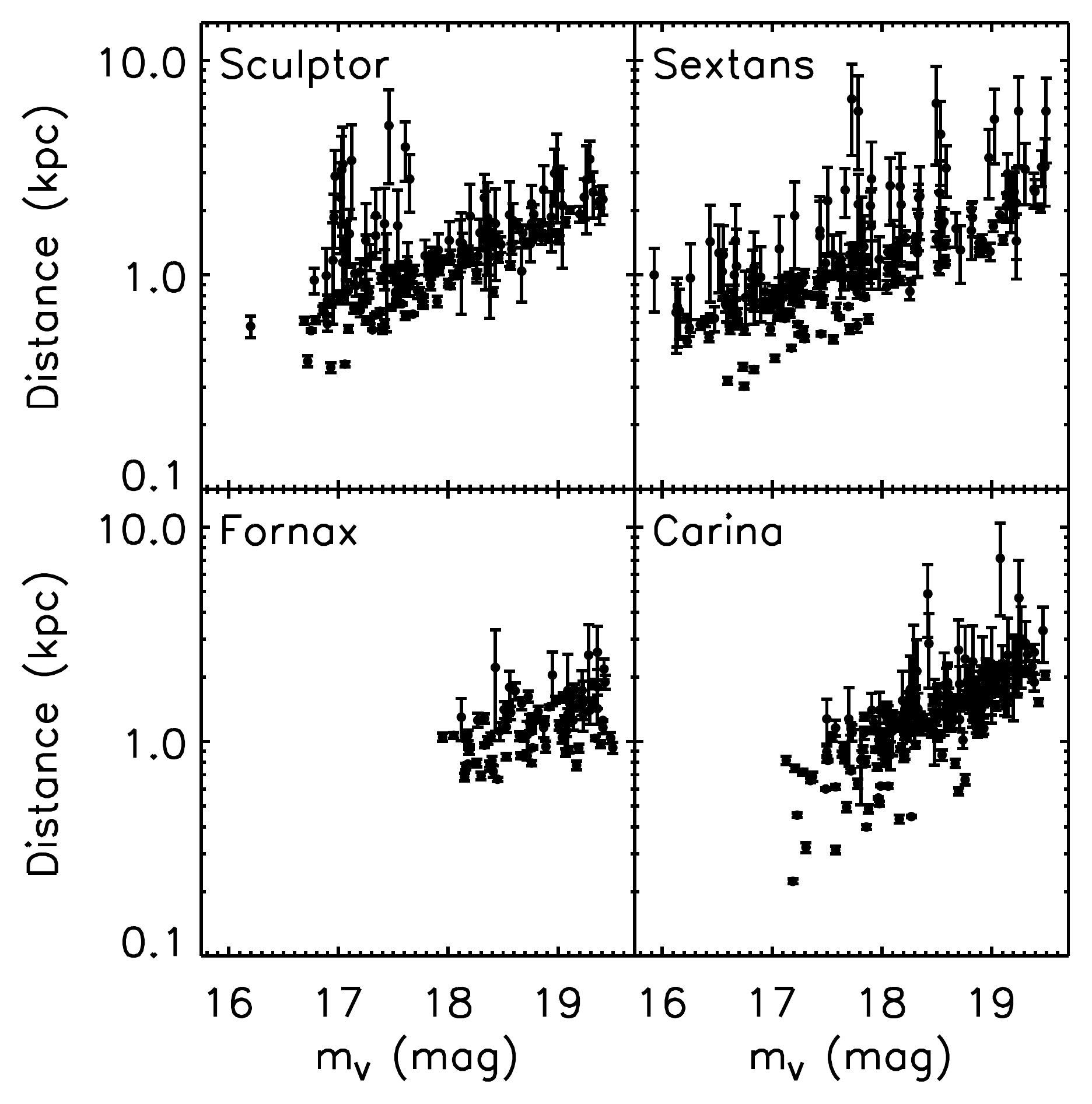}
   \caption{ V--magnitude versus derived line-of-sight distance for the foreground stars, as derived by our pipeline (in a logarithmic scale). The errors in the distances are as indicated. }%
   \label{Fig:Dist_errors}
    \end{figure}

Figure~\ref{Fig:Dist_errors} shows the derived distances versus the apparent V-magnitudes for the foreground stars (selected as in Sect.~\ref{sect:foregound_selection}), with individual error bars for each measurement (see Table~\ref{tab:other_parameters}, available electronically, for the individual values). Most of the stars follow a similar trend, within the error bars, with the faintest stars being also the most distant. This is consistent with a sample
dominated by dwarf stars, as expected for this study of the foreground sample.

Finally,  we verified our distance estimation pipeline by selecting the dSph targets (see Sect.~\ref{sect:foregound_selection}) for which good quality spectra were available (SNR$>$20~pixel$^{-1}$). The mean derived distances  are $74\pm17$~kpc, $60\pm21$~kpc, $117\pm18$~kpc and $109\pm27$~kpc, respectively for the dSph of Sculptor, Sextans, Fornax and Carina. Although these extragalactic stars have $\alpha-$abundances different from the ones assumed  in the grid of synthetic spectra \citep[see, for example][]{Tolstoy09}, and that for giant stars the uncertainties are greater than for the dwarfs for both the atmospheric parameters and the distances, we note that these values are in a fairly good agreement with the commonly admitted values of $72\pm5$~kpc, $83\pm9$~kpc, $120\pm8$~kpc and $85\pm5$~kpc for Sculptor, Sextans, Fornax and Carina, respectively \citep{Irwin95}.

\onltab{4}{
\begin{table*}
\caption{Positions, radial velocities, magnitudes, and distances of the selected foreground stars}\label{tab:other_parameters}
\begin{tabular}{lcccccccccc}
\hline \hline
ID & RA & dec & $m_V$ & $(V-I)$ & \vrad & $\Delta$\vrad & $D$ & $\Delta D$\\ 
      &  (deg)  & (deg) & (mag) & (mag) & (\kms) & (\kms) & (pc) & (pc) \\ \hline
             \multicolumn{5}{l} {line-of-sight: Sculptor        }  & \multicolumn{4}{r} {Foreground stars=  166                   } \\ \hline
       I-1    &   14.3761    &  -33.7909    &  18.20    &   0.96    &   19.10    &    4.07    &   1878    &    769   \\
       I-2    &   15.6097    &  -33.8622    &  18.01    &   1.12    &   58.07    &    4.07    &   1446    &    327   \\
       ...\\
        \hline \multicolumn{5}{l} {line-of-sight: Sextans        }  & \multicolumn{4}{r} {Foreground stars=  219                  } \\ \hline 
      II-1    &  153.1806    &   -1.5348    &  17.81    &   1.14    &   56.48    &    6.67    &   1131    &     68   \\
      II-2    &  153.2692    &   -1.7572    &  17.44    &   0.95    &   37.14    &    6.67    &   1608    &    710   \\
      ...\\
        \hline \multicolumn{5}{l} {line-of-sight: Fornax        }  & \multicolumn{4}{r} {Foreground stars=   86                  } \\ \hline 
     III-1    &   40.3831    &  -33.8863    &  18.56    &   1.07    &   -3.80    &    2.08    &   1791    &    341   \\
     III-2    &   40.3736    &  -33.8517    &  18.39    &   1.40    &   13.23    &    2.08    &   1058    &     40   \\
     ... \\
       \hline \multicolumn{5}{l} {line-of-sight: Carina        }  & \multicolumn{4}{r} {Foreground stars=  208                  } \\ \hline 
      IV-1    &  100.5212    &  -50.8089    &  19.01    &   1.07    &   20.95    &    3.15    &   2178    &    154   \\
      IV-2    &  100.5859    &  -50.8368    &  19.03    &   1.05    &   18.28    &    3.15    &   1829    &    135   \\
... \\ \hline
\end{tabular}
\tablefoot{Full table available  at the CDS.}
\end{table*}
}

\subsection{Identification of the foreground stars}
\label{sect:foregound_selection}

Foreground stars were isolated by the removal from the sample of all 'extragalactic stars', namely those stars with a projected \logg$<3.75$~cm~s$^{-2}$ (in order to select the giants), and having radial velocities  within $3\sigma$ of the nominal centre-of-mass radial velocity of the respective dSph in each line-of-sight (see Fig.~\ref{Fig:Vrad_vs_meta}). In this procedure, we adopted  the values of $(54\pm 11.4)$~\kms, $(110\pm10.1)$~\kms, $(226 \pm 8.4)$~\kms,
$(224 \pm 7.5)$~\kms\ as central and dispersion values of the radial velocity distributions of Fornax, Sculptor, Sextans and Carina, respectively \citep{Battaglia06,Battaglia08b, Battaglia11, Koch06}.

Furthermore, we removed all the spectra with radial velocity errors greater than 10~\kms\ ($\sim$2/3 of a sampled pixel), as determined in \cite{Battaglia08b}. Indeed, \citet{Kordopatis11a}, have shown that the pipeline's errors on the final atmospheric parameters, using spectra from the same setup, are not affected as long as the errors on the Doppler corrections are lower than this threshold value.  We also removed all the stars for which the error on the distance was higher than 50\%, in order to obtain more robust results in the following analysis. It should be noted that such a threshold on the distance error removes preferentially turn-off stars, since the projection of the pipeline's parameters on the isochrones for stars in  that evolutionary phase will result to a higher uncertainty in the derivation of their absolute magnitude.

Finally, we made an additional selection based on the apparent magnitude of the stars, discarding the faintest targets, with $V>$19.5~mag.  We made a cut on the received flux (apparent magnitude), rather than on SNR,  in order to avoid the possibility of introducing spurious selection biases due to our definition of SNR. 
Recall that, as in \cite{Kordopatis11a}, the mean SNR per pixel of a given spectrum is obtained as follows:  given a solution spectrum template, we compute for each pixel the difference with the observation, and then select only the adjacent pixels with a flux close to 1, for which the difference changes sign. The SNR is then estimated by measuring the dispersion of the selected pixels.
This method has been shown in \cite{Zwitter08} to give relatively reliable measures of the
SNR, but  in the case of the most metal-poor stars, the dispersion in flux of the continuum pixels is usually over-estimated, due to a combination of  the facts that they are very numerous and  that for these stars there are very few  suitable spectral signatures for accurate determination of the atmospheric parameters.  Similarly, faint metal-rich stars are also measured with a lower SNR than in reality, since some of the metallic lines can be mis-identified as noise.
That being said, we did discard any star having SNR$<2$~pixel$^{-1}$ since this low a value indicates that either the stellar spectrum is missing or that the fibre was not been centred properly on the star (so that the received flux did not scale correctly with apparent magnitude).


Applying the above rejection criteria, we selected \Nforeground\ foreground stars in total: \Nscl\ in the Sculptor line-of-sight, \Nsxt\ towards the Sextans line-of-sight, \Nfnx\ towards the Fornax line-of-sight and \Ncar\ towards the Carina line-of-sight.
The stars span distances from 200~pc up to  $\sim 7$~kpc, with the majority of the targets closer than 3~kpc from the Galactic plane. 

  \begin{figure}[t]
   \centering
\includegraphics[width=0.5\textwidth]{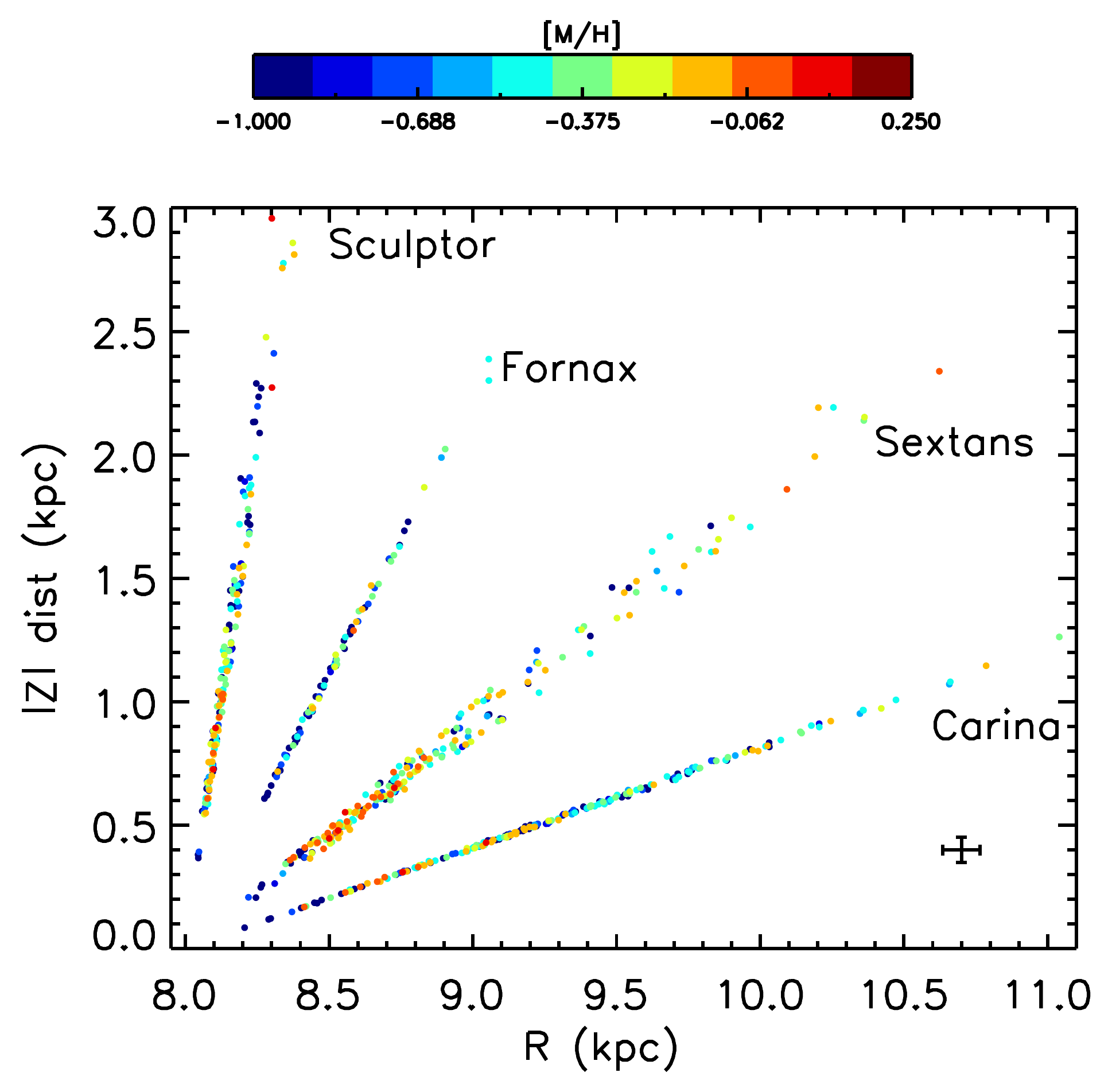}
   \caption{ Absolute distance above the Galactic plane (Z) versus the Galactocentric radial distance ($R_\odot=8$~kpc) for the foreground stars. The colour code goes from blue for the most metal-poor stars to red for the most metal-rich. The pencil beams, from left to right, correspond to the lines-of-sight of Sculptor, Fornax, Sextans and Carina, respectively. The median error for the entire foreground dataset is represented at the lower right corner.}%
   \label{Fig:Z_vs_R}
    \end{figure}

 Figure~\ref{Fig:Z_vs_R} shows the position of the stars in the $ |Z| $ versus R plane, with the points being colour-coded  according to the  stellar metallicity. We assumed that the Sun is located at $R_\odot=8$~kpc \citep{Reid93}. As may be seen from the figure, the mean metallicity of the stars separates into two regimes, with the division at around $\sim 0.8$~kpc from the Galactic plane. Closer than $|Z|\sim0.8$~kpc, the stars are predominantly metal-rich ([M/H]$>-0.5$~dex), whereas farther than 0.8~kpc, they are mostly metal-poor ([M/H]$<-0.5$~dex). The metallicity and  $|Z|$ distance of this division suggest that we are seeing the transition from predominantly thin disc stars to  thick disc stars. Indeed, with a scale-height of $h_Z\sim$800-1000~pc, compared to $h_Z \sim$200-300~pc for the scale-height of the thin disc, the distance above the Galactic plane at which the thick disc is expected to become the dominant population is around 1~kpc, given a local density of approximately 10--15\% for the thick disc stars \citep[see for example][and references therein]{Soubiran03, Girard06, Juric08, deJong10}.



We decided to discard the Fornax line-of-sight from the rest of the analysis, given the relatively small number of foreground stars identified towards this direction ($N=$\Nfnx), as well as  the relatively small range in distances spanned by these stars.
 In addition, the radial velocity of the Fornax dSph lies within the range of the foreground disc stars, which could result in some background stars (dSph members) contaminating the foreground sample if their surface gravity were incorrectly estimated.

\subsection{Estimation of the azimuthal orbital velocity $V_\phi$}

No proper motions are available for the foreground stars of our sample and hence no 3D motions can be derived. Nevertheless, the radial velocity, combined with the distance estimation, can give us valuable information about the azimuthal motion of the stars ($V_\phi$)  which in turn can be used to classify stars as belonging to either the thick or thin disc -- or to some other component of the Galaxy. 
For this purpose we follow the approach of \cite{Morrison90,Wyse06}.

We define   $V_{Gal}$ as the Galactocentric velocity of a star along the line-of-sight from the Sun to the star as: 
\begin{equation}
V_{Gal}=V_{\rm rad} + v_{{\rm pec},\odot,(l,b)} + V_{LSR}\sin l \cos b
    \end{equation}
    where \vrad~is the heliocentric radial velocity (line-of-sight velocity),     $v_{{\rm pec},\odot,(l,b)}=15.4$~\kms~ is the adopted peculiar motion of the Sun in the direction $(l,b)=(51^{\circ},+23^{\circ})$, and $V_{LSR}=220$~\kms~ is the adopted circular velocity of the local standard of rest.

    In spherical coordinates, one can also decompose $V_{Gal}=\alpha V_r +\beta V_\phi + \gamma V_\theta$, where the coefficients depend on the position of the stars in the Galaxy, and can be expressed as a function of the Galactic coordinates $(l,b)$ and $D$, the line-of-sight distance. Given the reasonable assumption that the mean motions in $r-$ and $\theta-$ are zero, $\hat{V_\phi}=V_{Gal}/\beta$ can be used as an unbiased estimator of the true azimuthal velocity $V_\phi$.

The coefficient $\beta$ may be expressed in terms of the projection of the heliocentric distance $D$ on the Galactic plane ($u$) and the distance of the Sun from the Galactic centre  ($R_\odot=8$~kpc):
\begin{eqnarray}
u^2 &=& R_\odot^2+D^2 \cos^2 b - 2DR_\odot \cos l \cos b \\
\beta &=& \frac{1}{u} R_\odot \cos b \sin l 
       \end{eqnarray}

          \begin{figure}[t]
   \centering
\includegraphics[width=0.45\textwidth]{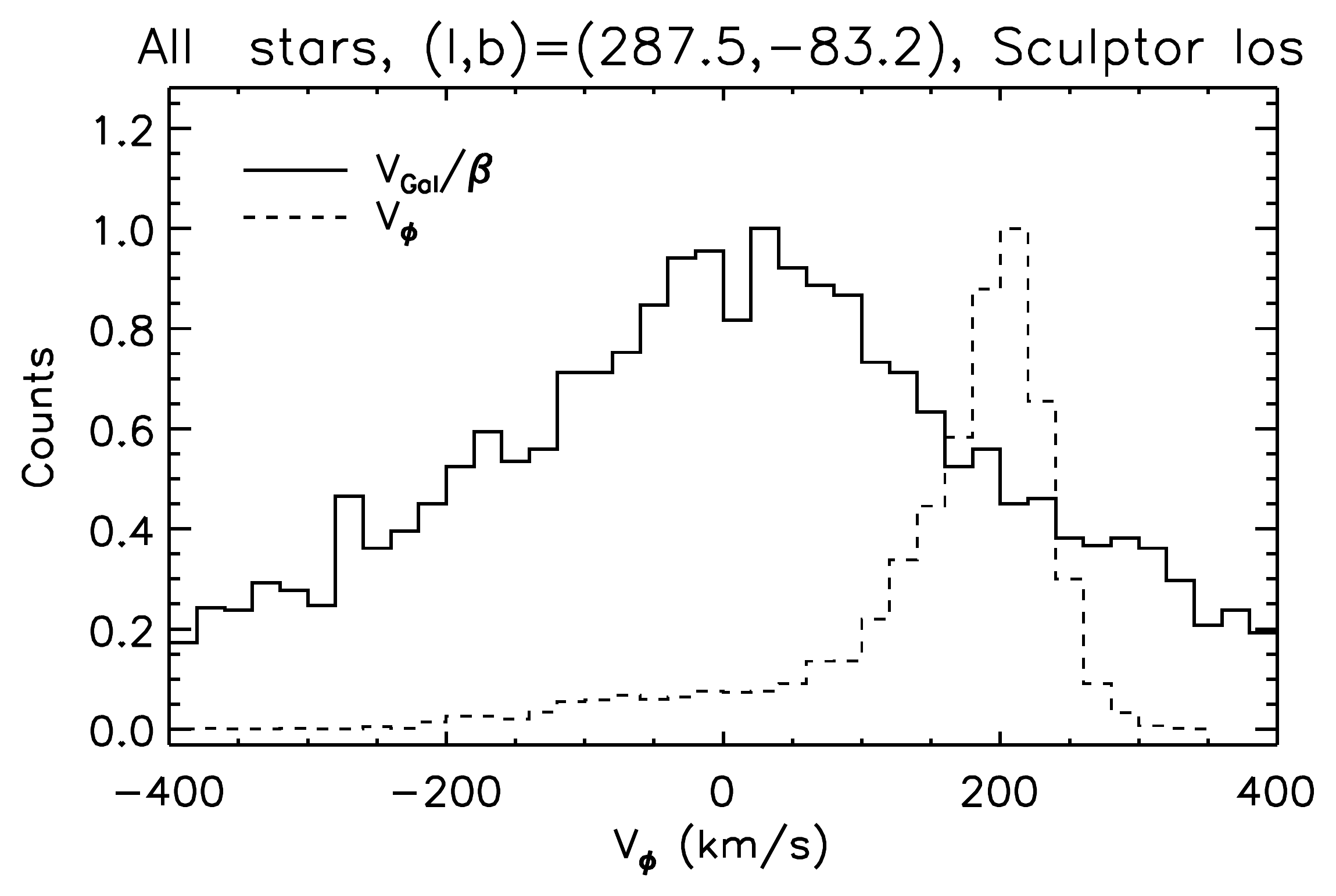}\\
\includegraphics[width=0.45\textwidth]{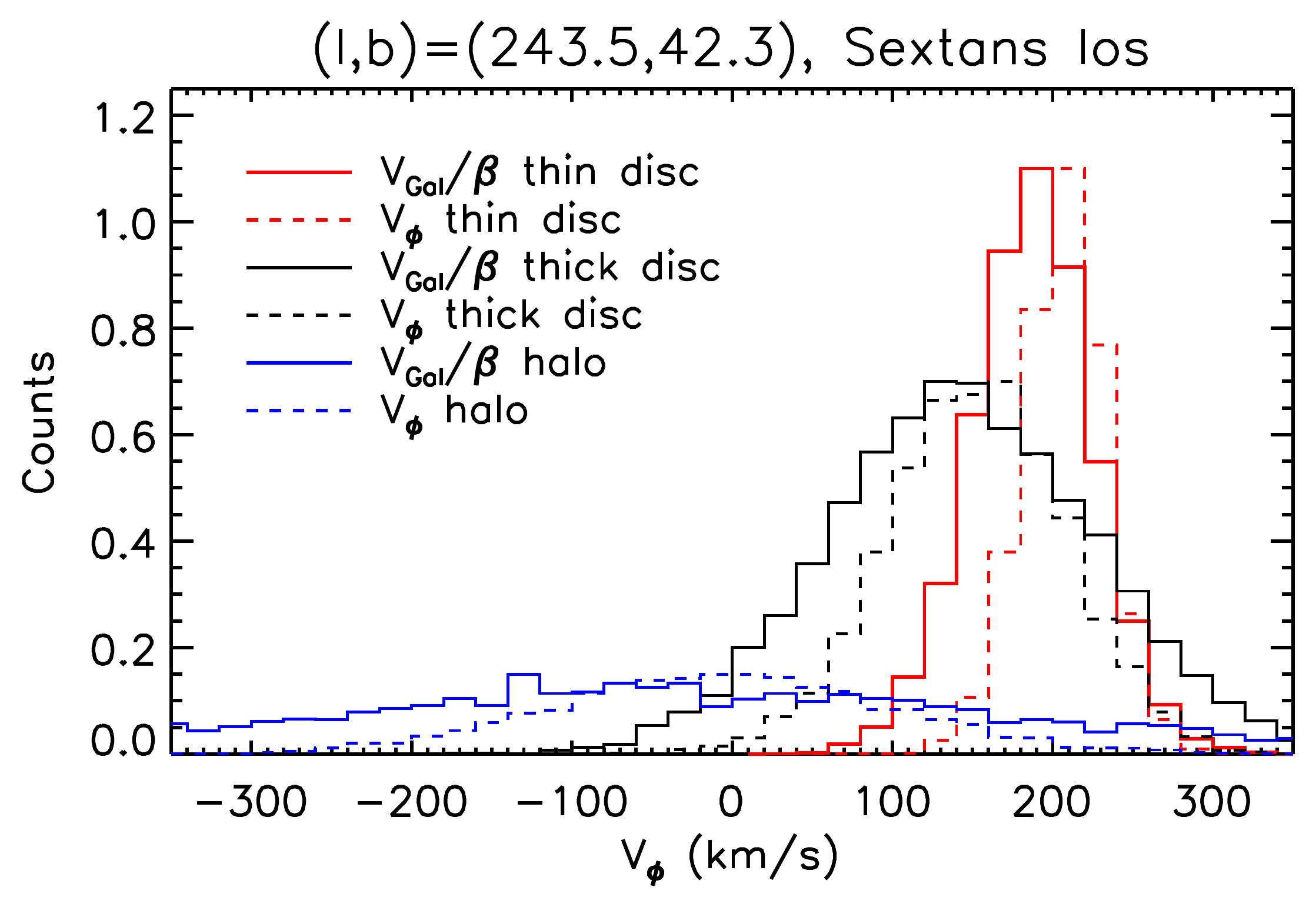}\\
\includegraphics[width=0.45\textwidth]{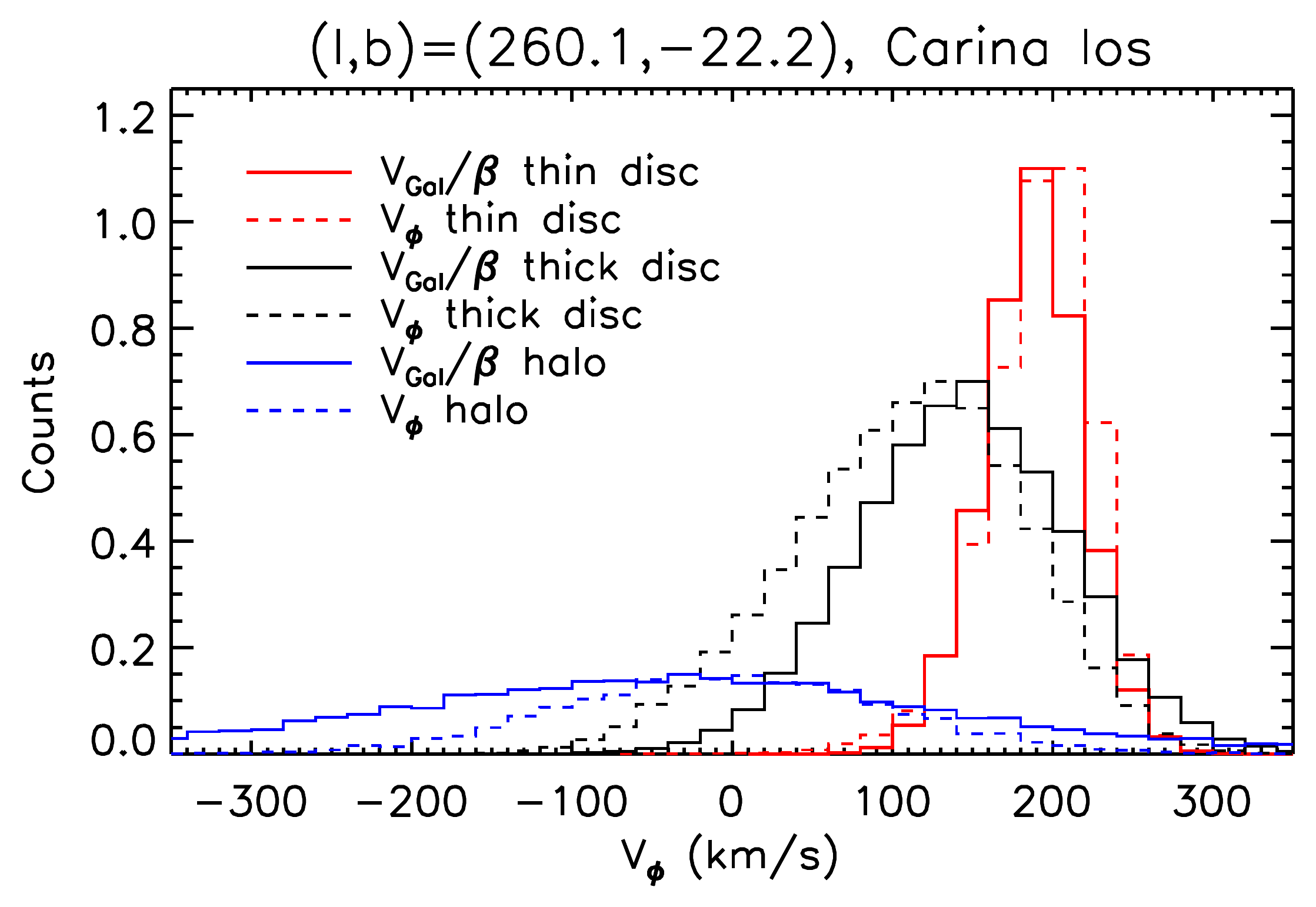}
\caption{True azimuthal velocity $V_\phi$ (dashed histograms) computed from simulations of the Besan\c{c}on model and the estimator $\hat{V_\phi}=V_{Gal}/\beta$ (solid  histograms) computed from the same simulations for the lines-of-sight of Sculptor (top), Sextans (middle) and Carina (bottom). For the lines-of-sight of Sextans and Carina, the velocity histograms have been decomposed into thin disc (red), thick disc (black) and halo (blue). The normalisation of each Galactic population is arbitrary, and has been chosen for reasons of clarity.}
   \label{Fig:Vphi_estim}
    \end{figure}

  For stars close to the tangent point ($l=270^\circ$), as is the case for these dSph lines-of-sight, the above equations show that $\hat{V_\phi}$ will be a good estimator of $V_\phi$ at low Galactic latitudes, whereas the closer the line-of-sight is to the poles, the less the contribution of \vrad~ to $V_\phi$ (since $\beta \sim 0$). This is illustrated in Fig.~\ref{Fig:Vphi_estim}, which shows histograms of both the $\hat{V_\phi}$ estimator and the true $V_\phi$ for three different lines-of-sight.  The distributions of values for these plots have been obtained from the predicted stellar positions, velocities and proper motions of the Besan\c{c}on model (see next section). It can be seen that while  for the Sextans and Carina line-of-sight, $\hat{V_\phi} \sim V_\phi$, this is clearly not the case for the Sculptor line-of-sight, where the estimator for the azimuthal velocity  cannot be used for the purpose of the present analysis.  This is not too surprising, since for this high-latitude line-of-sight \vrad $\sim - V_Z$.

\section{ Characterisation of the chemical properties of the Galactic discs}
\label{Sect:Results}

\subsection{Creation of a comparison catalogue with the Besan\c{c}on Galactic model}

\begin{figure}[tbp]
\begin{center}
$\begin{array}{c}
      \includegraphics[width=0.33\textwidth]{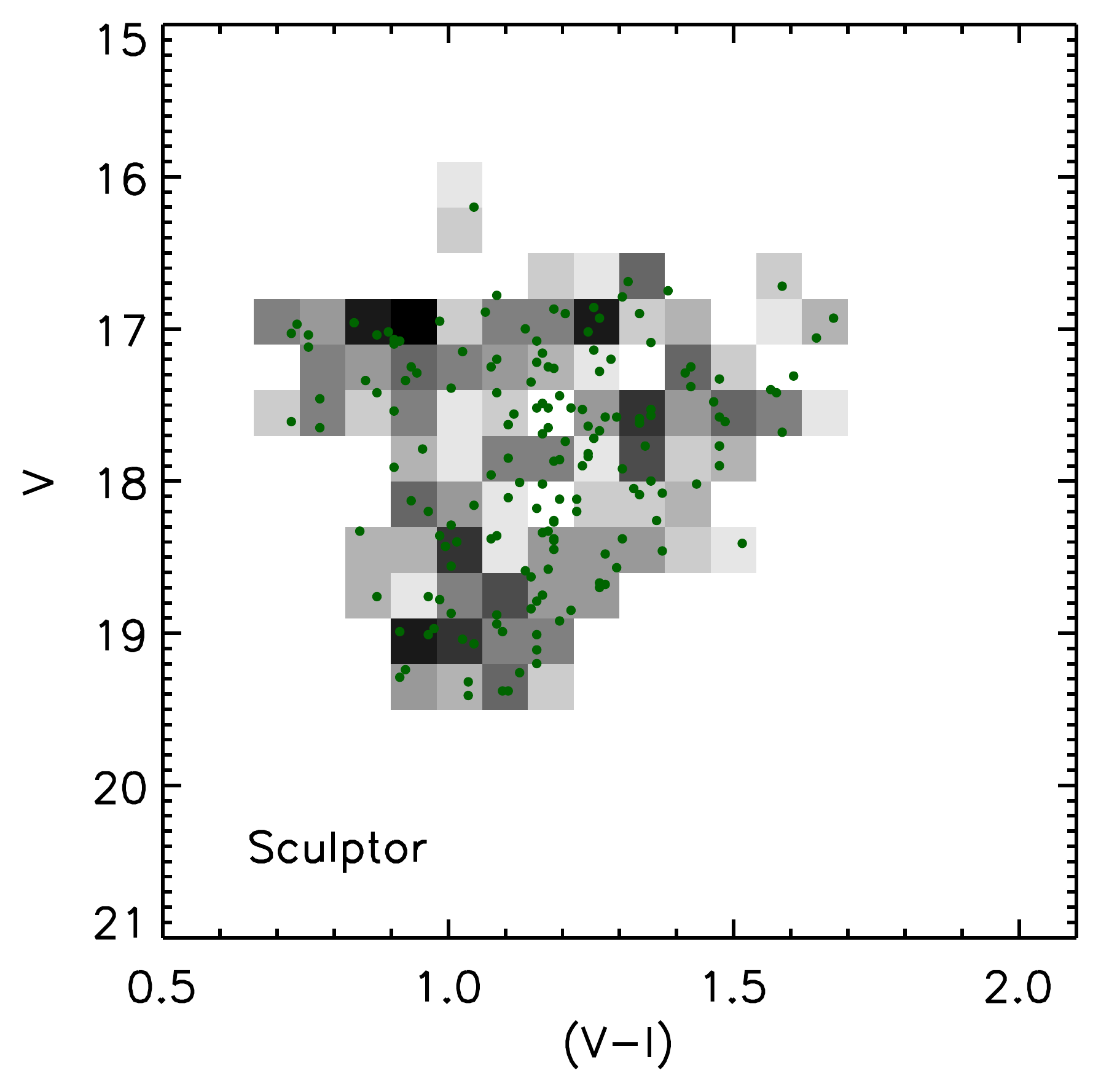} \\   \includegraphics[width=0.33\textwidth]{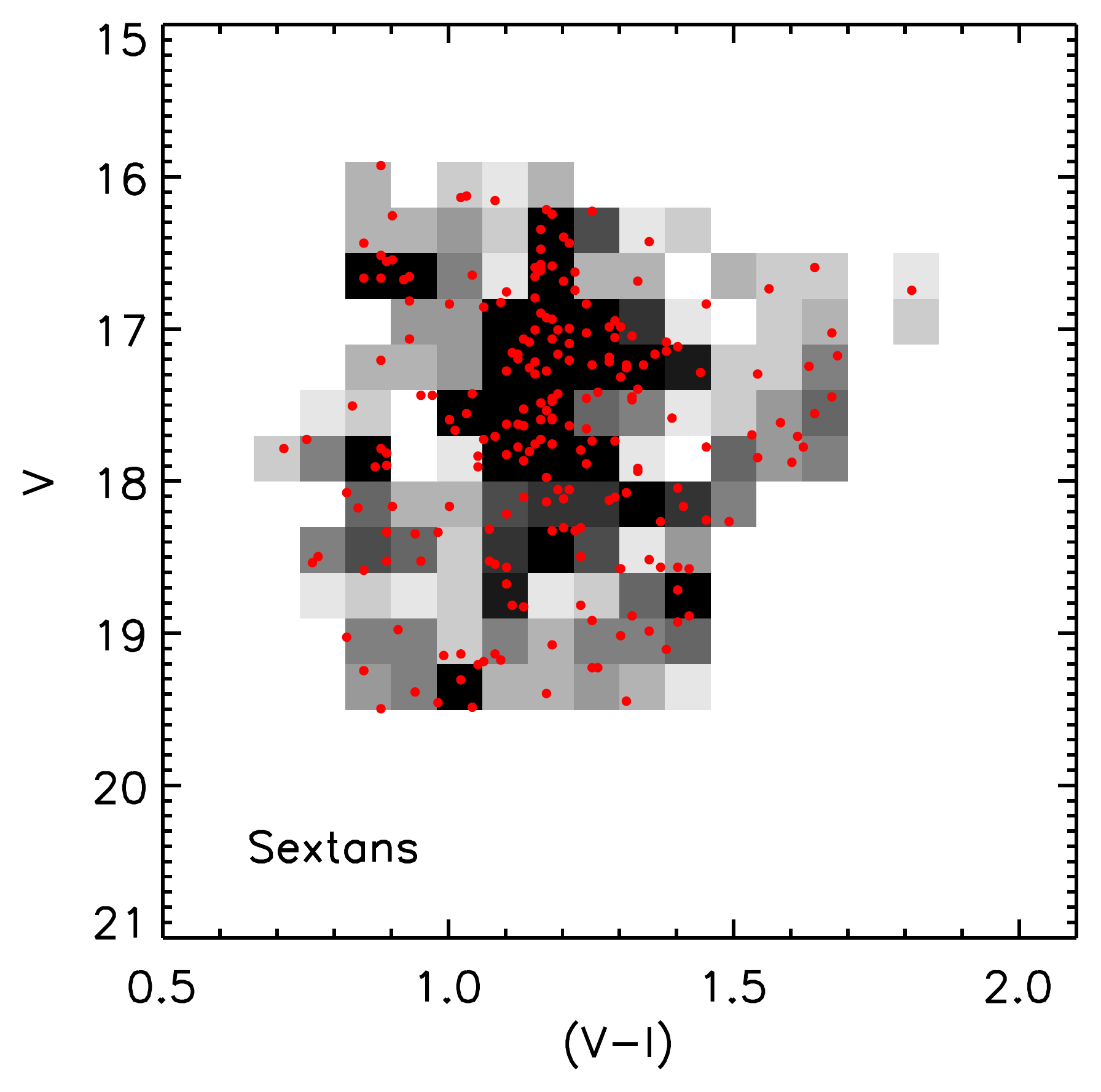} \\   \includegraphics[width=0.33\textwidth]{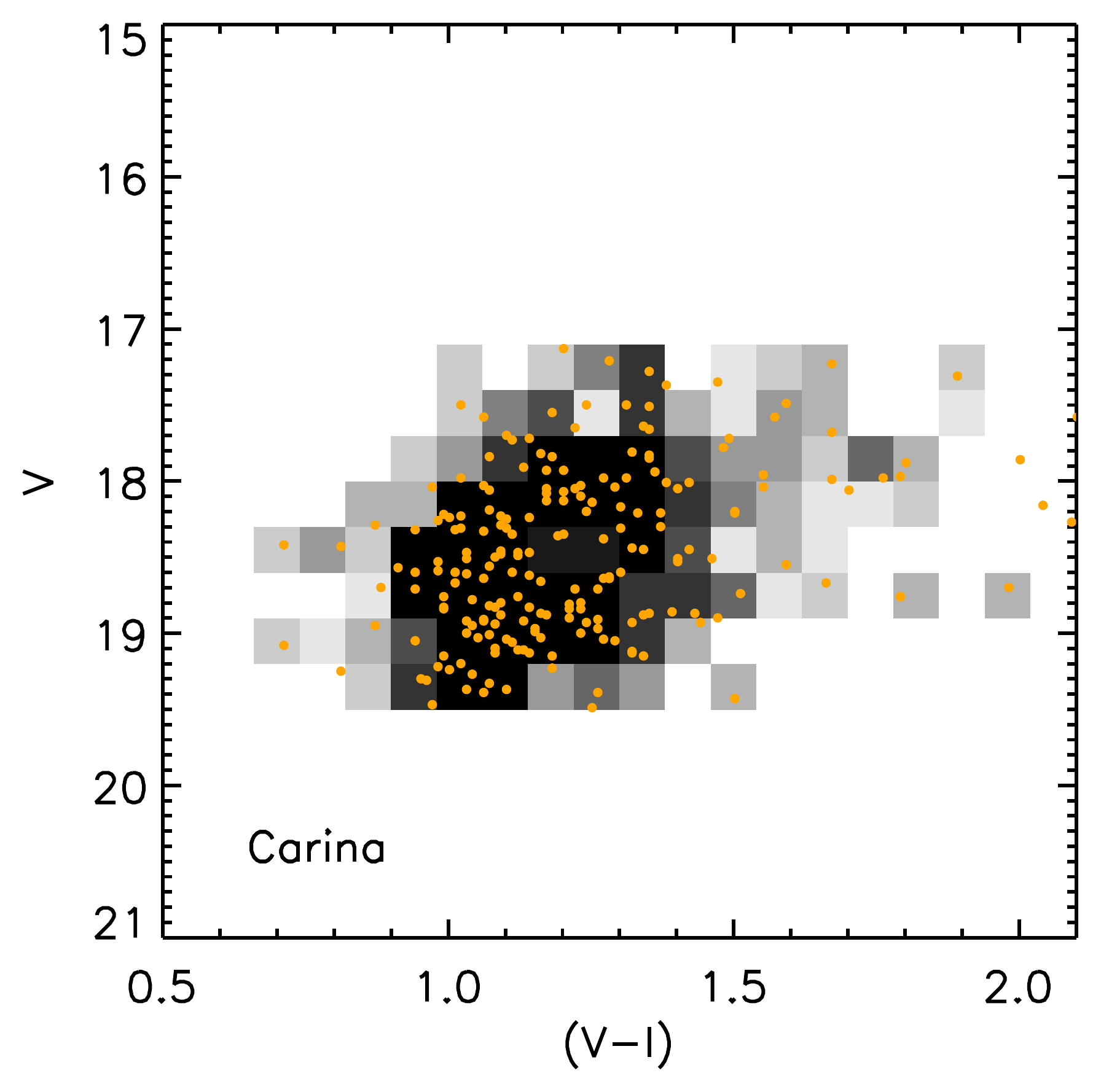} 
		\end{array}$
\end{center}
\caption{$V$--magnitude versus $(V-I)$ colour for the foreground data (coloured circles) and the biased Besan\c{c}on models (shown in grey-scale in each panel). The simulated samples follow the same $V$ and $(V-I)$ distributions as the selected observed stars.  }
   \label{Fig:Bias_models}
\end{figure}

The interpretation of our results is facilitated by comparisons with simulated populations in each line-of-sight obtained from the Besan\c{c}on model of the Galaxy\footnote{http://model.obs-besancon.fr/} \citep{Robin03}.  We recall that in the Besan\c{c}on model, 
 the Galactic thin disc has  both vertical and radial metallicity gradients, the vertical one being driven by the superposition of seven stellar populations of different ages and metallicities, and the radial one being  fixed at $\partial$[Fe/H]/$\partial R=-0.07$~dex~kpc$^{-1}$.
On the contrary, there are no metallicity gradients within the modelled Galactic thick disc, which is assumed to have a mean metallicity of $-0.78$~dex,  lower compared to the values of $\sim-0.6$~dex, as  derived by most recent investigations \citep[see, for example,][]{Bensby07, Kordopatis11b}. We therefore adjusted the mean metallicity of the simulated thick-disc population by the addition of 0.2~dex.  The simulated stellar catalogues were then used to select  sub-samples to match our observations, through the imposition of the same $V-$magnitude and $(V-I)$ distributions as in the data (see Fig.~\ref{Fig:Bias_models}). 
In addition, the simulated Galactic giant stars (\logg$<3.75$~cm~s$^{-2}$) that had a radial velocity within the distribution of the dSph of that line-of-sight were removed.  Finally, for each simulated star we assigned typical errors on the metallicities and distances, matching the general trend in errors of our observations. Thus for each metallicity and distance value in the simulated catalogues, we defined a Gaussian error distribution function of standard deviation $\sigma_D$ or $\sigma_{\rm [M/H]}$, from which a random error value was drawn and then assigned to the given parameter. The standard deviations that were assumed for the distances and the metallicities were $\sigma_D = 0.07 \times D$ and $\sigma_{\rm [M/H]}=-0.12 \times {\rm [M/H]} +0.45$~dex, respectively.  We note that the error modelling overestimates errors for the good quality spectra (e.g. those of high SNR),  but underestimates them for the low quality ones. Nevertheless, the present analysis does not require a more sophisticated model for the errors.

\subsection{The Sculptor and Sextans lines-of-sight}

 \begin{figure*}[tbph]
\begin{center}
$\begin{array}{cc}
      \includegraphics[width=0.45\textwidth,angle=180]{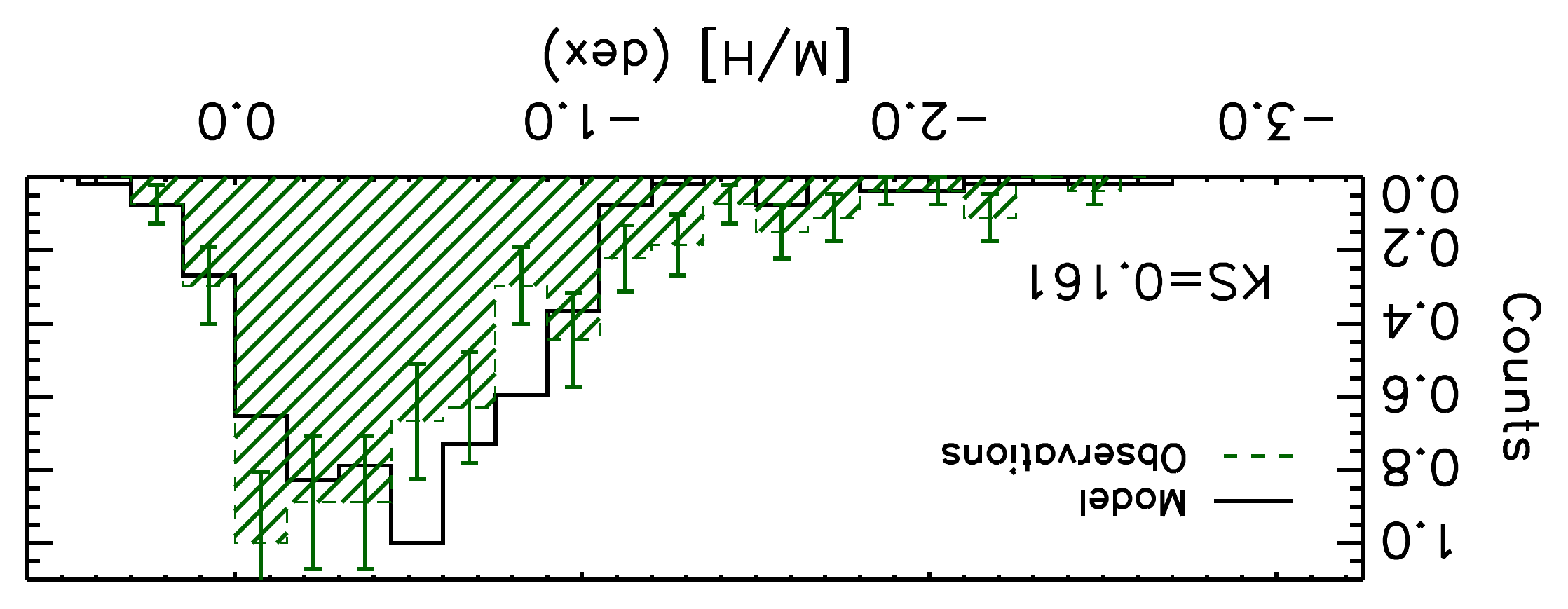} & \includegraphics[width=0.45\textwidth,angle=180]{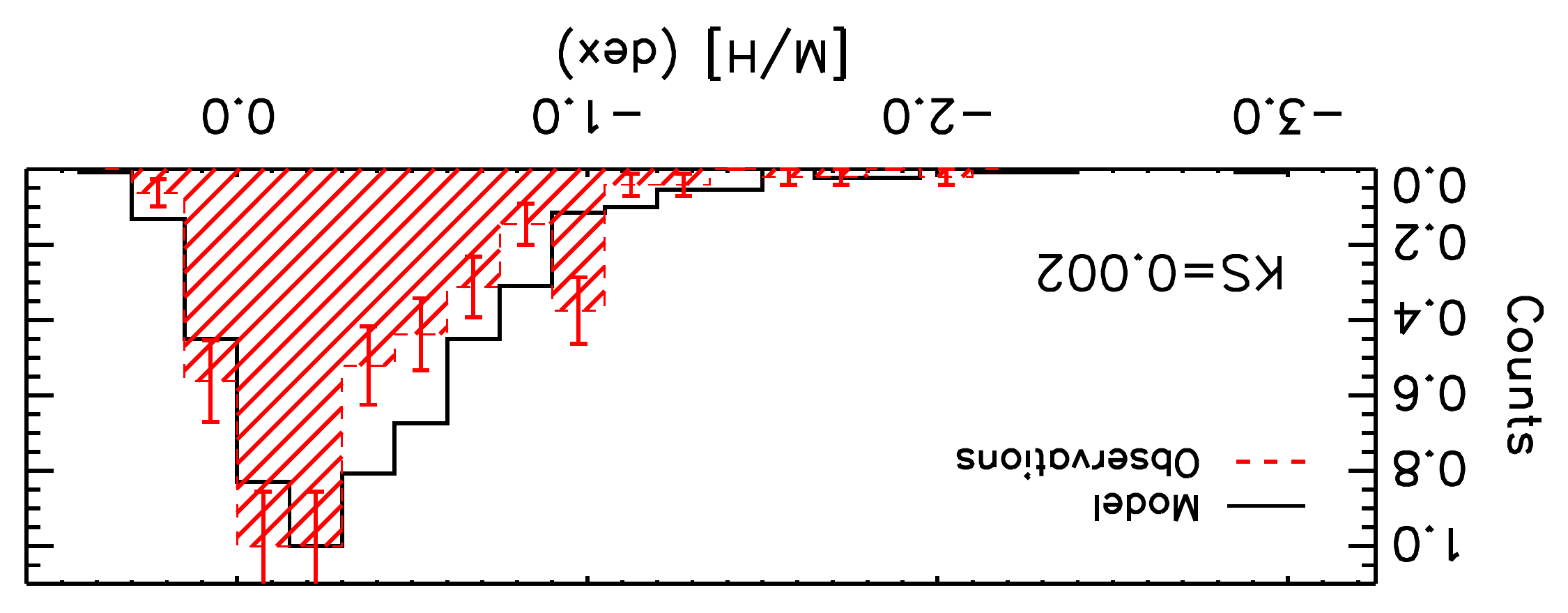}\\
      	\includegraphics[width=0.45\textwidth]{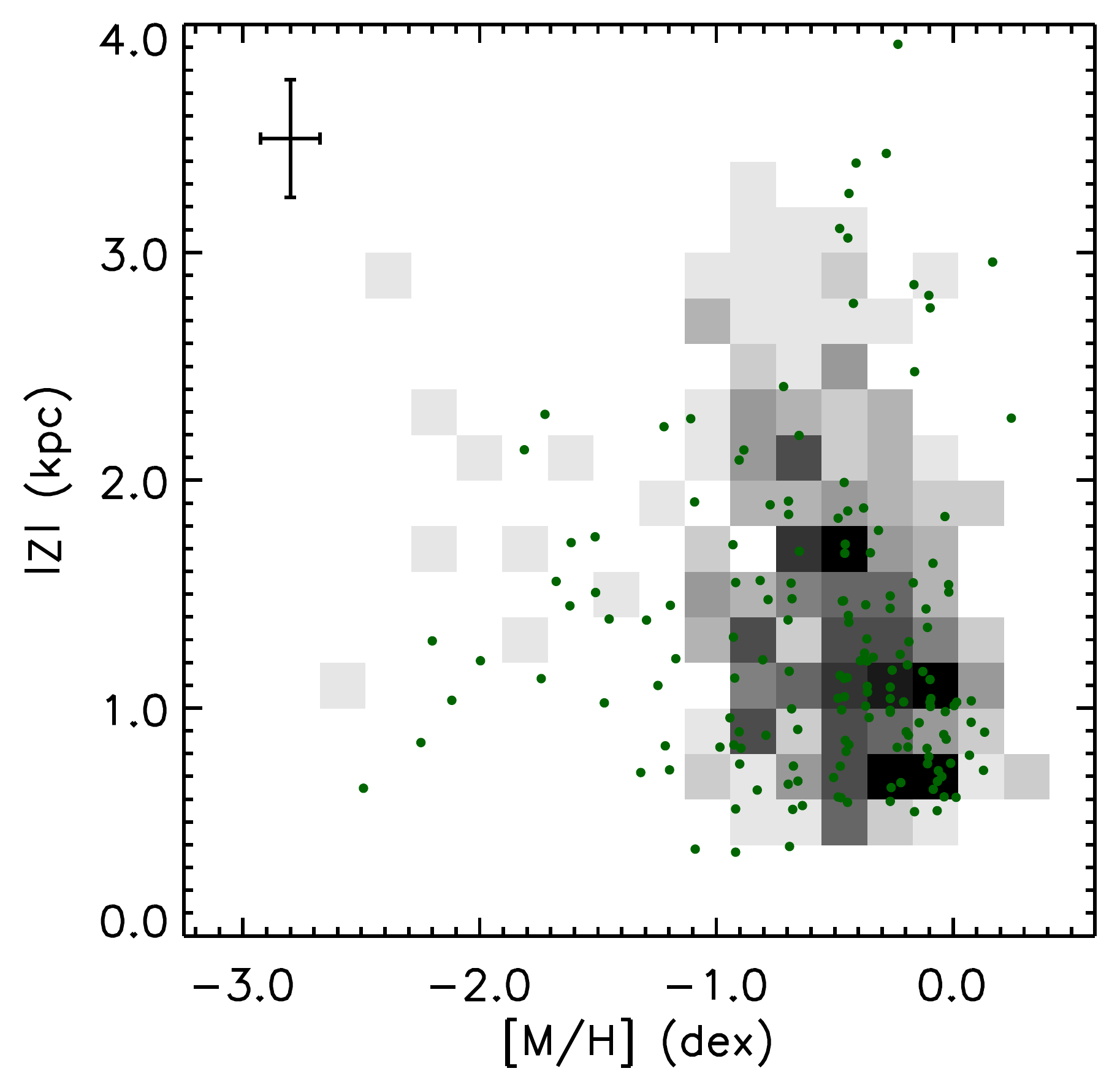} & \includegraphics[width=0.45\textwidth]{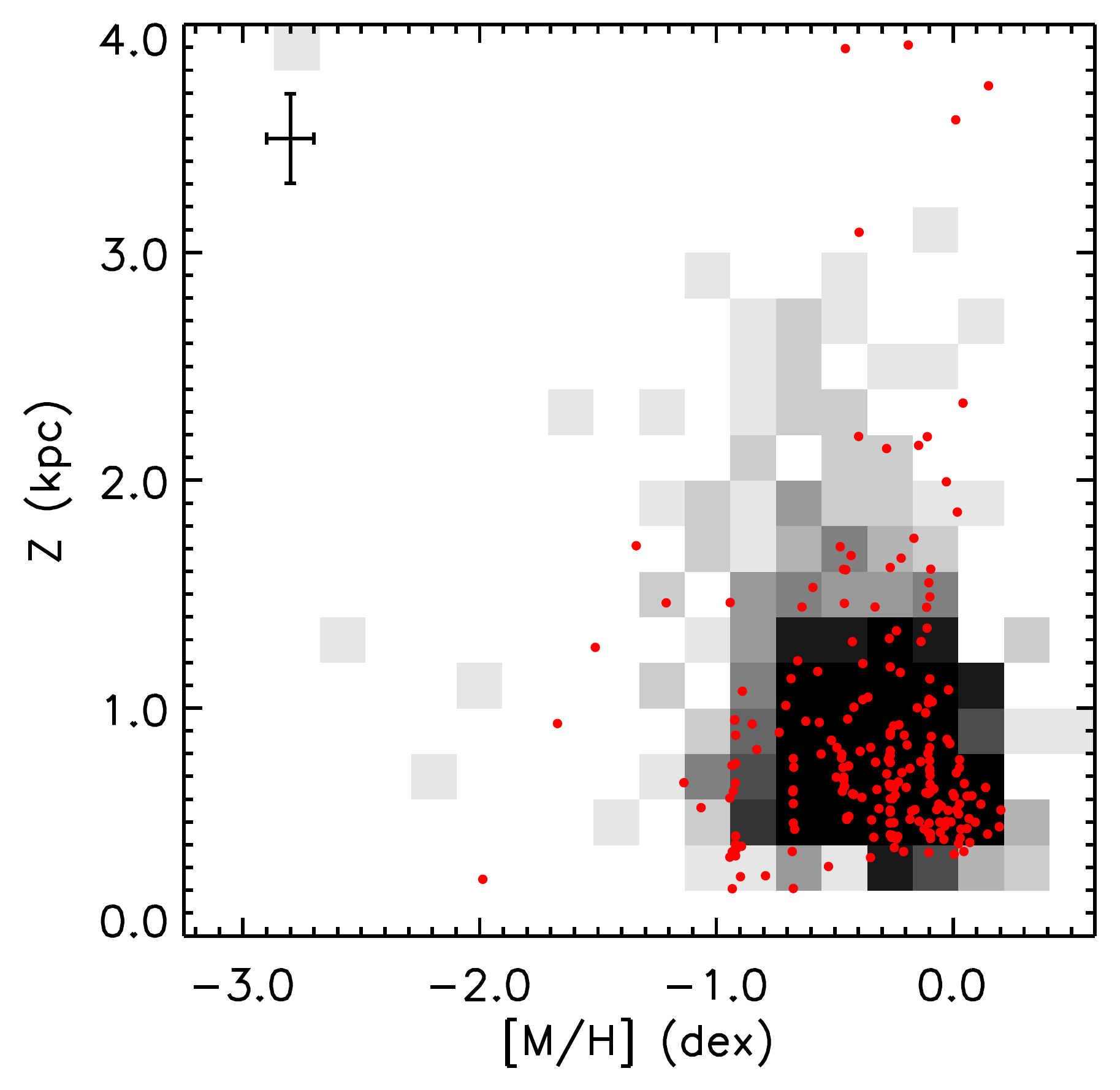}  \\
	       	\includegraphics[width=0.462\textwidth]{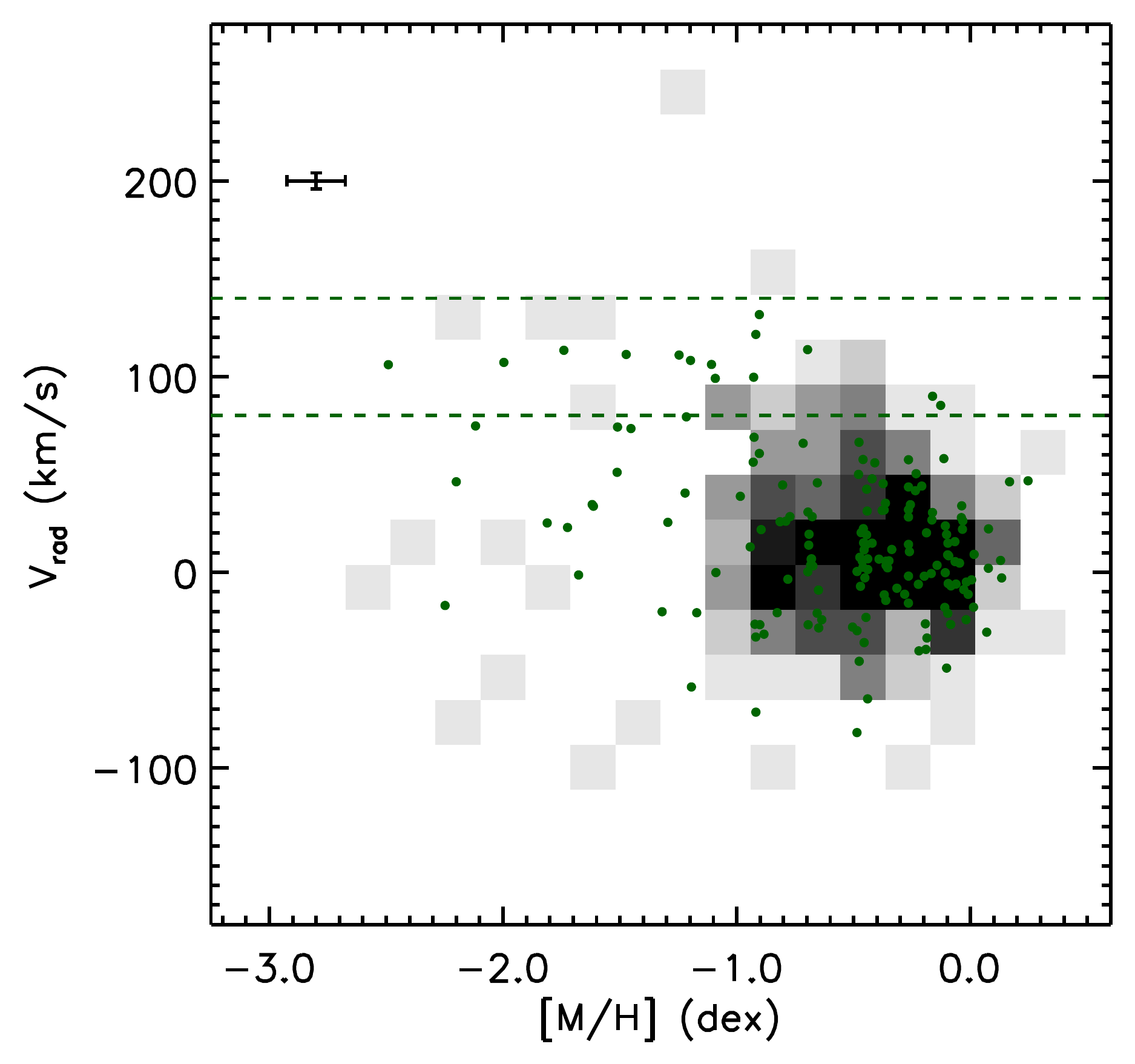}  & \includegraphics[width=0.462\textwidth]{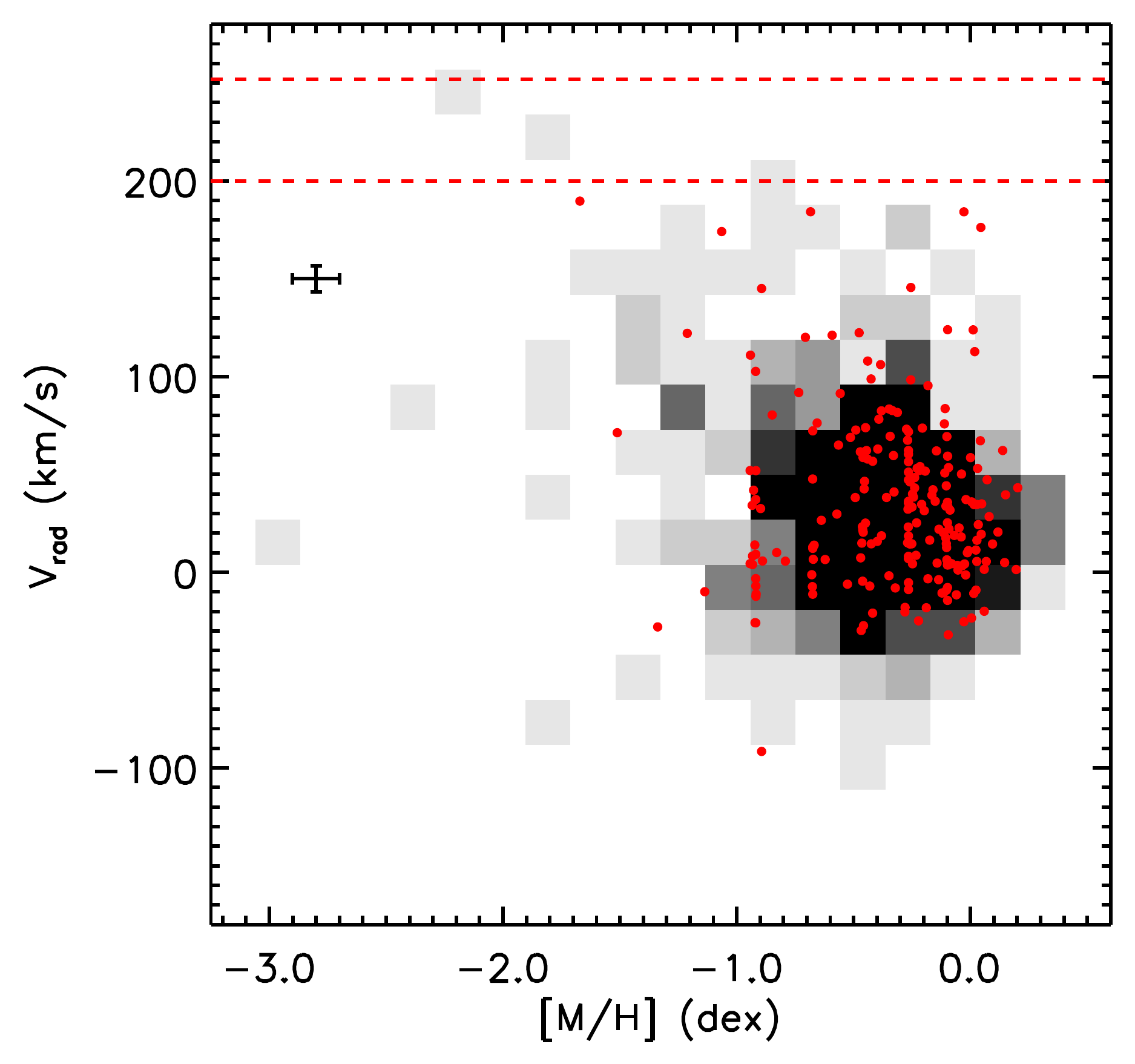} \\
		\end{array}$
\end{center}
\caption{Position above the Galactic plane (middle panels) and radial velocity (bottom panels) versus metallicity for the observations (in green for the Sculptor line-of-sight on the left-hand side and red for the Sextans line-of-sight on the right-hand side). The predictions of the Besan\c{c}on model are shown in grey-scale.  The horizontal dashed lines  indicate the range of radial velocities appropriate for member stars of the respective dSph. The panels at the very top show the metallicity distributions of the foreground stars in these two lines-of-sight and the Kolmogorov-Smirnov probability of not rejecting the null hypothesis.}
   \label{Fig:Comp_Scl_Sxt_Bes}
\end{figure*}
    
\begin{figure*}[th]
   \begin{center}
   $\begin{array}{cc}
\includegraphics[width=0.5\textwidth,angle=180]{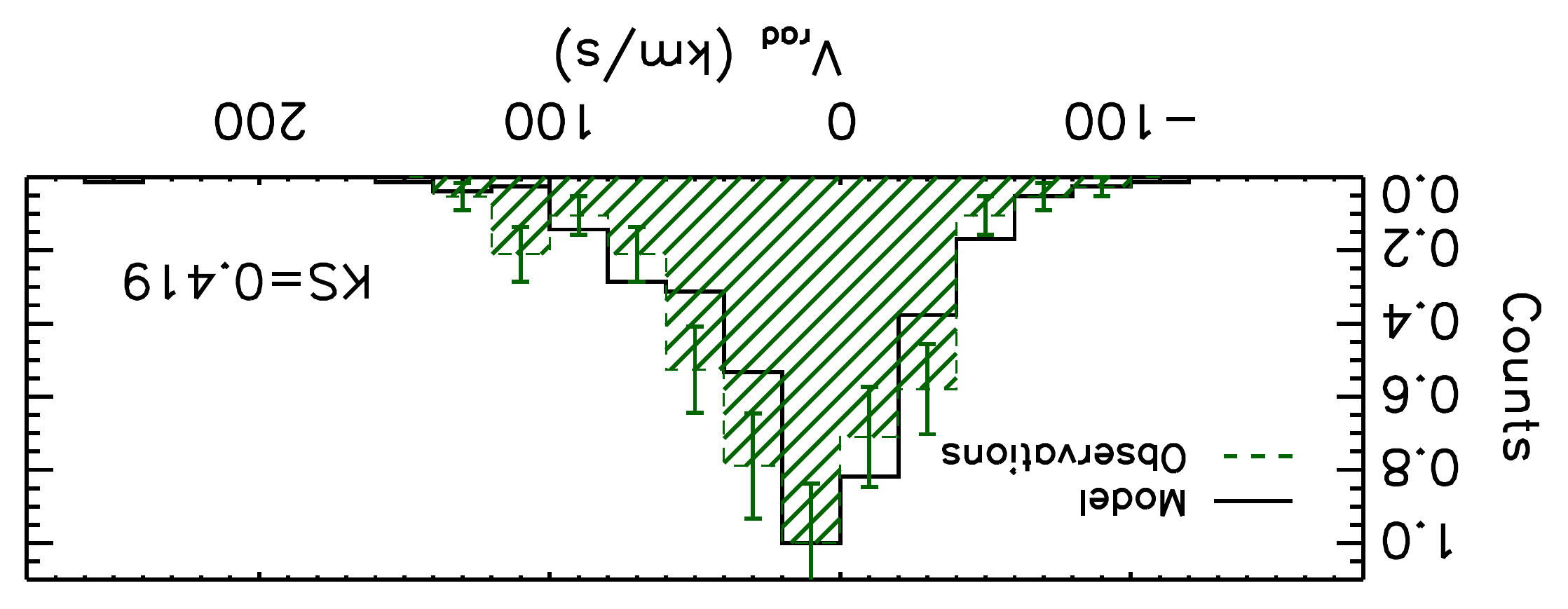} & \includegraphics[width=0.5\textwidth,angle=180]{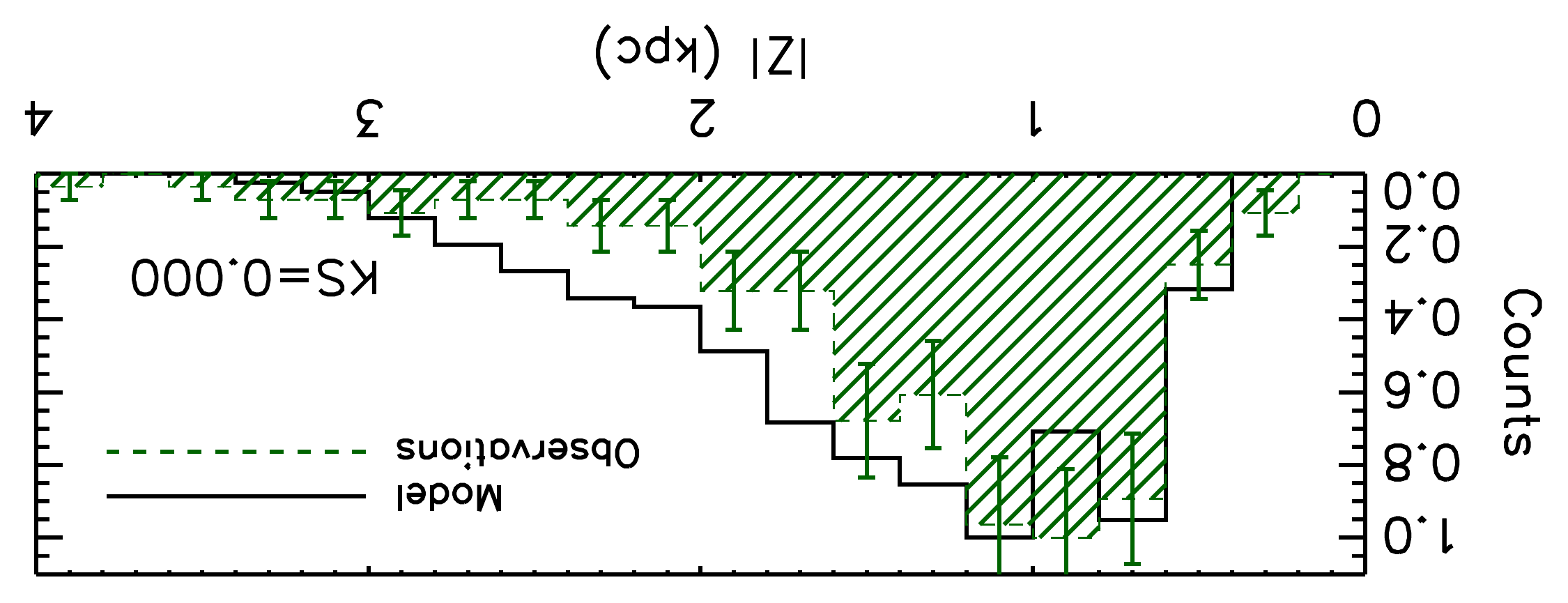}  \\
\includegraphics[width=0.5\textwidth,angle=180]{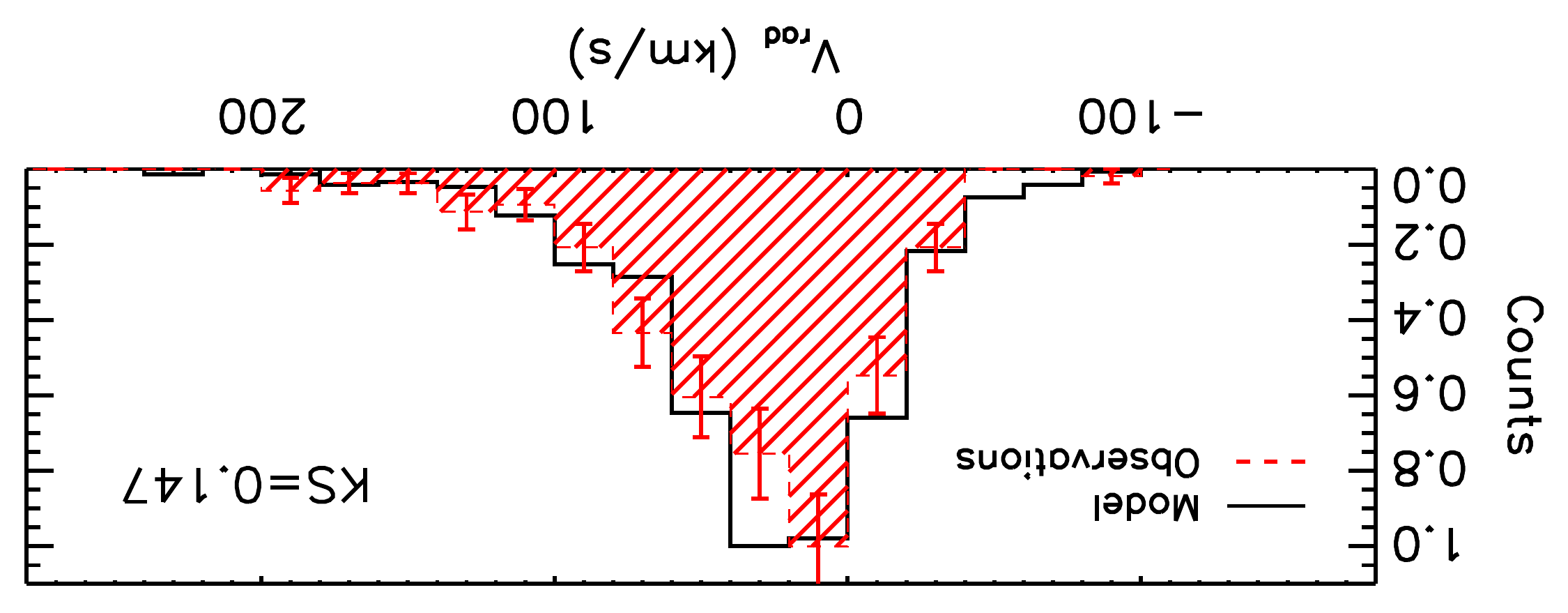} & \includegraphics[width=0.5\textwidth,angle=180]{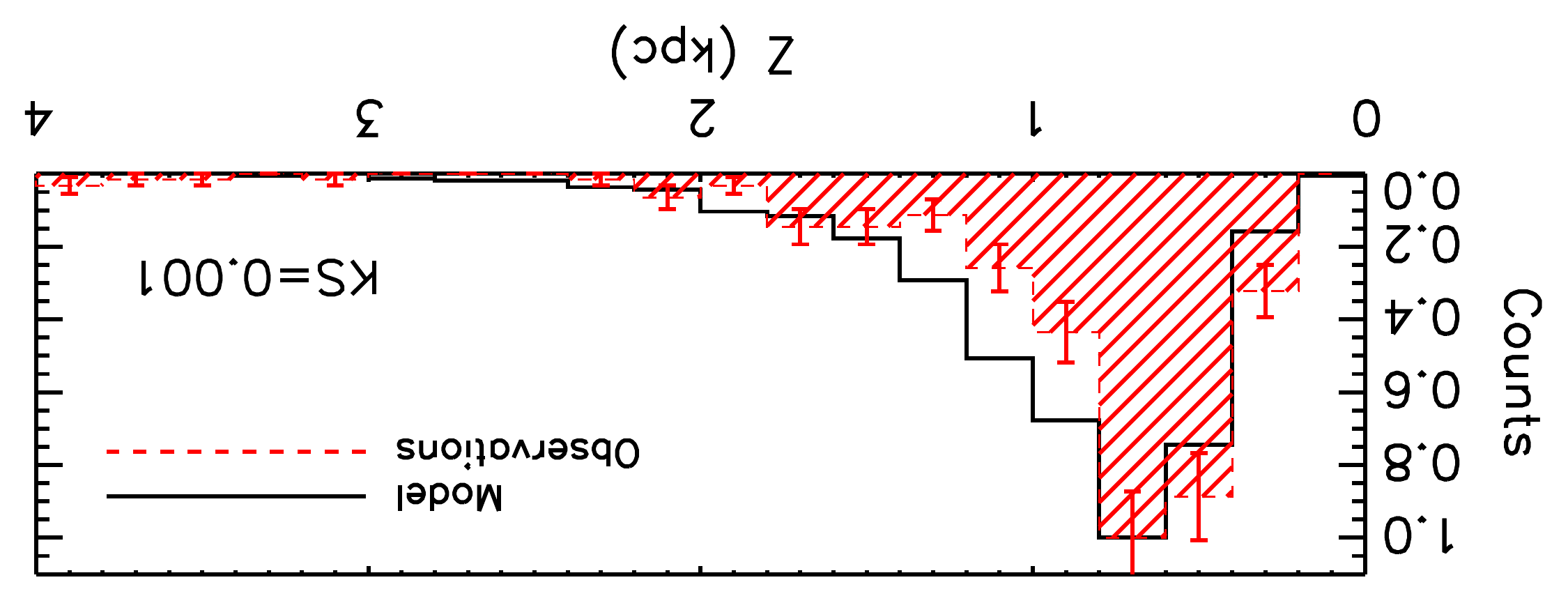} \\
		\end{array}$
\end{center}
   \caption{Comparisons of the histograms of the simulated catalogue and the observations towards the line-of-sight of Sculptor (upper panel) and the line-of-sight of Sextans (lower panel). The Kolmogorov-Smirnov probability of not rejecting the null hypothesis is written in the right side of each histogram, and is representative of a single Monte-Carlo realisation.   }%
   \label{Fig:Scl_comparisons}
    \end{figure*}

 Figure~\ref{Fig:Comp_Scl_Sxt_Bes} shows Z-distance from the Galactic plane and radial velocities versus metallicity for the observed  foreground stars, together with the predictions from the Besan\c{c}on models (after appropriate photometric selection to match that of the surveys).  
A first remark that can be made is that the model predicts more stars far from the plane (see Fig.~\ref{Fig:Scl_comparisons}), and that the most distant stars in the actual dataset are metal-rich. This is a consequence of our adoption of a rejection criterion based on the relative error of the distance -- by keeping only the targets with errors lower than 50\% we inevitably reject the faintest, metal-poor stars. Given the fact that mostly dwarf foreground stars are observed, the rejected targets are hence also the most distant.

 Furthermore, in the direction of the Sculptor line-of-sight, the comparison between the model and the data  shows an excess in the number of observed metal-poor stars at  $0.8 \leq |Z| \leq 2$~kpc, the typical distances where the thick disc dominates. These stars exhibit a radial velocity distribution with a range of roughly 100~\kms,  broader than one would expect for the canonical thick disc, though similar to the value expected in that line-of-sight for the stellar halo.  We note, however, that some of these stars might be contaminants from the Sculptor dSph,  wrongly assigned to the foreground by an  erroneous estimation  of surface gravity (as seen from their radial velocities in the bottom-left plot of Fig.~\ref{Fig:Comp_Scl_Sxt_Bes}).


We used Kolmogorov-Smirnov tests to compute the probabilities of the null hypotheses that the modelled radial velocities and metallicities are drawn from the same distributions as the observed ones.  As far as the radial velocities are concerned (see Fig.~\ref{Fig:Scl_comparisons}), the null hypothesis cannot be rejected for either  of the two lines-of-sight, with a probability constantly higher than $\sim 10\%$. In particular, in the case of the Sculptor line-of-sight (where \vrad $\sim - V_Z$), this probability can reach values higher than 50\%, depending on the Monte-Carlo realisations.  As far as the observed metallicities in the Sculptor line-of-sight are concerned, the null hypothesis that the simulation comes from the same underlying distribution as the observed one is valid with a probability higher than $\sim 10\%$. The agreement between the simulations and the observations for the Sextans line-of-sight is worse, with the probability of not rejecting the null hypothesis being less than 1\%.

   \begin{figure}[t]
   \centering
   \includegraphics[width=0.4\textwidth]{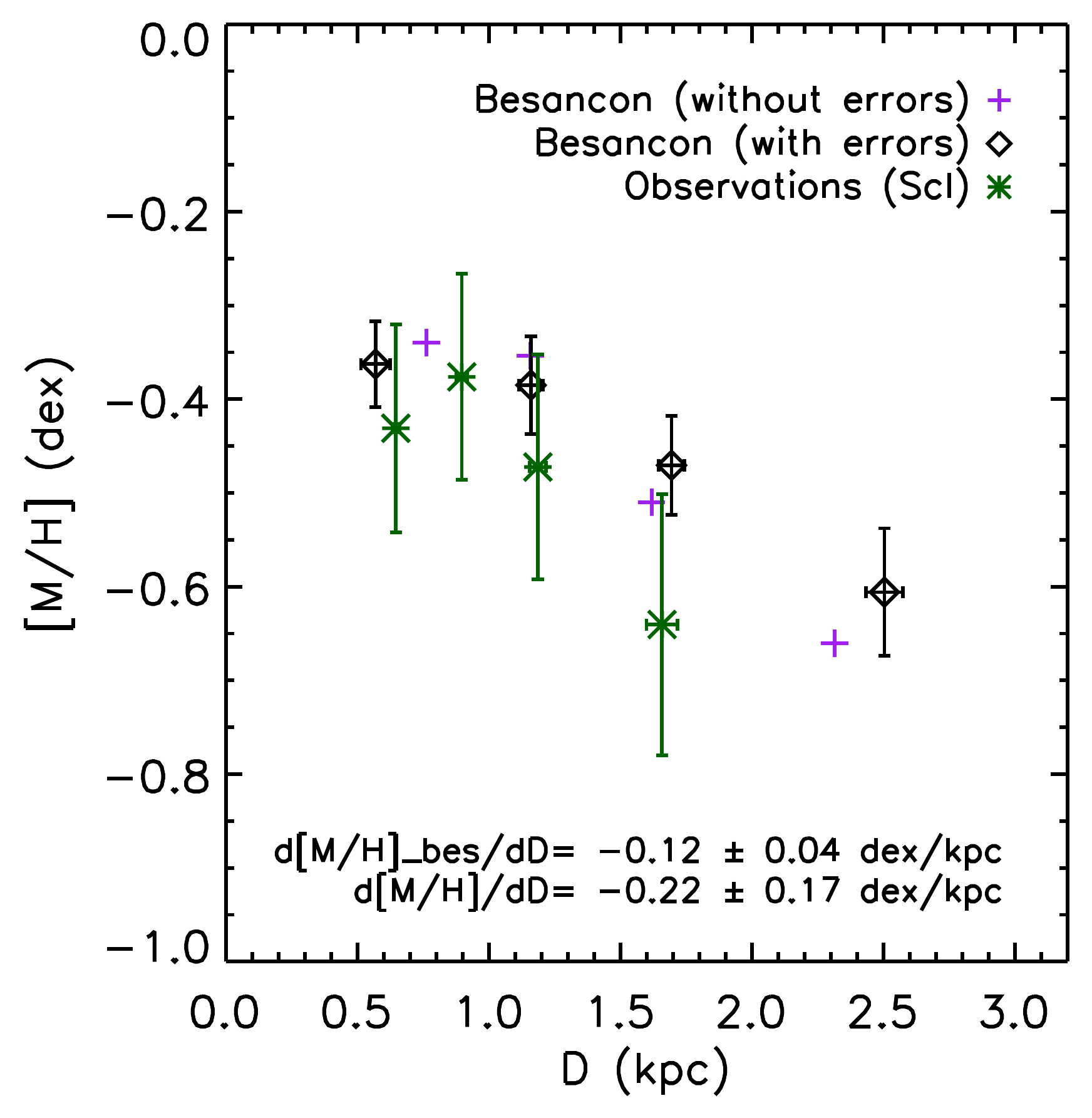}\\
\includegraphics[width=0.4\textwidth]{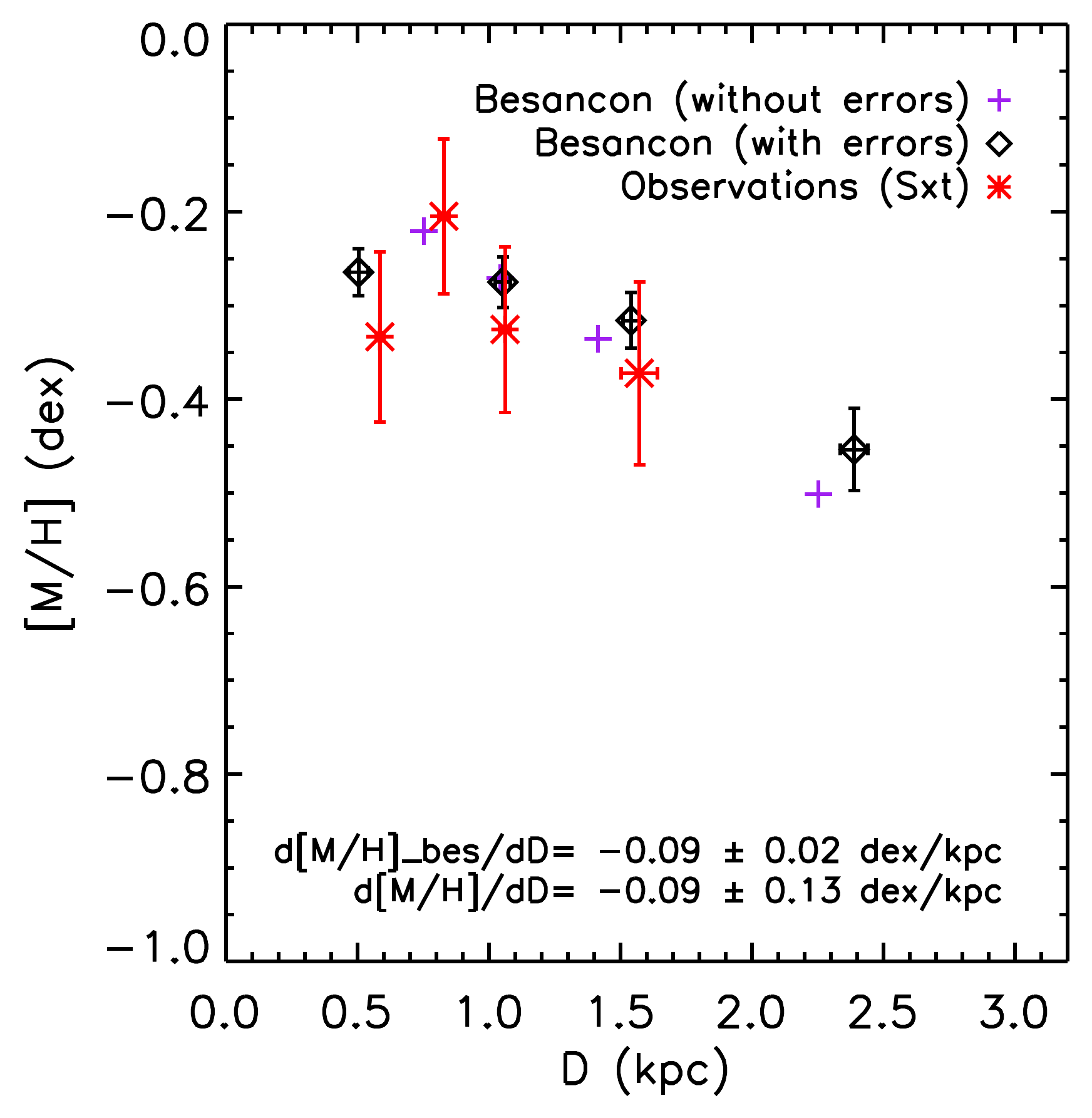}
   \caption{Observed and predicted (black diamond symbols) metallicity gradients for the Sculptor (top) and Sextans  (bottom) lines-of-sight. In each plot, the purple plus signs indicate the results obtained from the Besan\c{c}on model, assuming zero  errors on position and metallicity. }%
   \label{Fig:D_gradients}
    \end{figure}
    
    In order to interpret how smoothly the thin/thick disc (and eventually the thick disc/halo) transition happens along the different directions, we computed the observed and predicted mean metallicities at different line-of-sight distances, $D$.  Unfortunately, without any information  for either the 3D velocities or the $\alpha$-abundances, it is not possible to differentiate the different populations along the line-of-sight and hence investigate their individual intrinsic metallicity gradients.  We note, however, that the positions of the stars can provide insight into the nature of the surveyed population. In the case of the Sculptor line-of-sight, for example, where $D \sim Z$, the thin disc is expected to  dominate at distances closer than$\sim 0.7$~kpc, whereas towards the Sextans line-of-sight, the thin disc should still be dominant\footnote{For stellar populations with similar scale-length, the distance to which a given population will dominate star counts is proportional to $2 h_Z / \sin b$, where $h_Z$ is the scale-height of that population. }  up to $D\sim 1$~kpc.

We made $5\times 10^3$ Monte-Carlo realisations for both the distances and the metallicities, assuming that the errors for those two parameters are independent and Gaussian, centred on the derived values and with standard deviation equal to the associated uncertainties. For each realisation we separated the stars in  equally populated distance bins, and measured their median metallicity and distance.
Figure~\ref{Fig:D_gradients} shows the mean values of the medians of the Monte-Carlo realisations, the associated error bars being the dispersion of the median values across realisations.  Given these mean values and error bars, the metallicity gradients along the lines-of-sight have then been measured by fitting a slope using a least-mean-square method. 

We found that the metallicity gradients along both directions are negative, as expected since the lines-of-sight are pointing towards the outer  Galaxy and span relatively large distances far from the Galactic plane. The Sculptor line-of-sight has the steepest gradient, with \metaDgrad$=-0.22 \pm 0.17$~dex~kpc$^{-1}$, whereas towards the Sextans line-of-sight we found a gradient with amplitude  $-0.09 \pm 0.13$~~dex~kpc$^{-1}$.


For both lines-of-sight, the predictions (represented in the Fig.~\ref{Fig:D_gradients} by black diamond symbols) are in a fair agreement with the observations.  However, the simulations towards Sculptor show a slightly shallower metallicity gradient. This difference in this  line-of-sight indicates either that the relative importance of the thick disc is higher than that predicted, or that the thick disc has an intrinsic vertical metallicity gradient. Nevertheless, given the large error bars, any intrinsic gradient should be rather small, {\it i.e.}  less than $-0.10$~dex~kpc$^{-1}$.

\begin{figure}[t]
   \centering
\includegraphics[width=0.4\textwidth,angle=180]{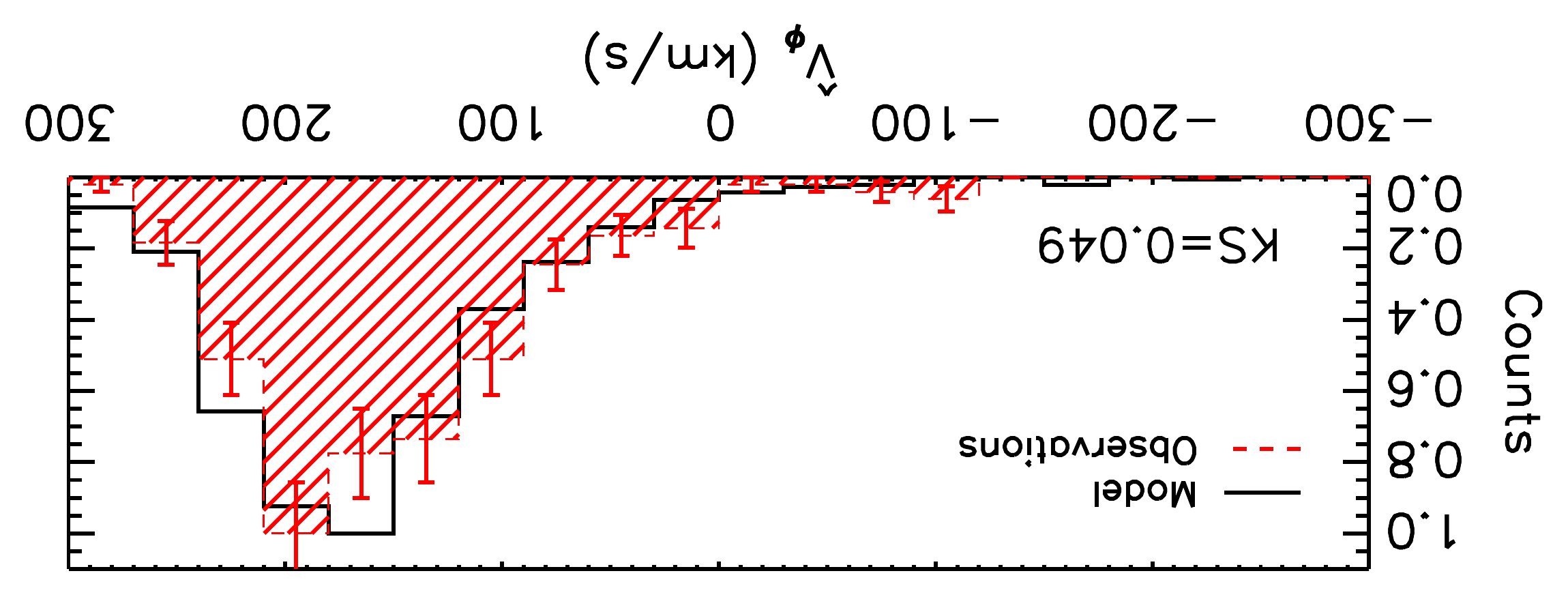} \\
\includegraphics[width=0.4\textwidth]{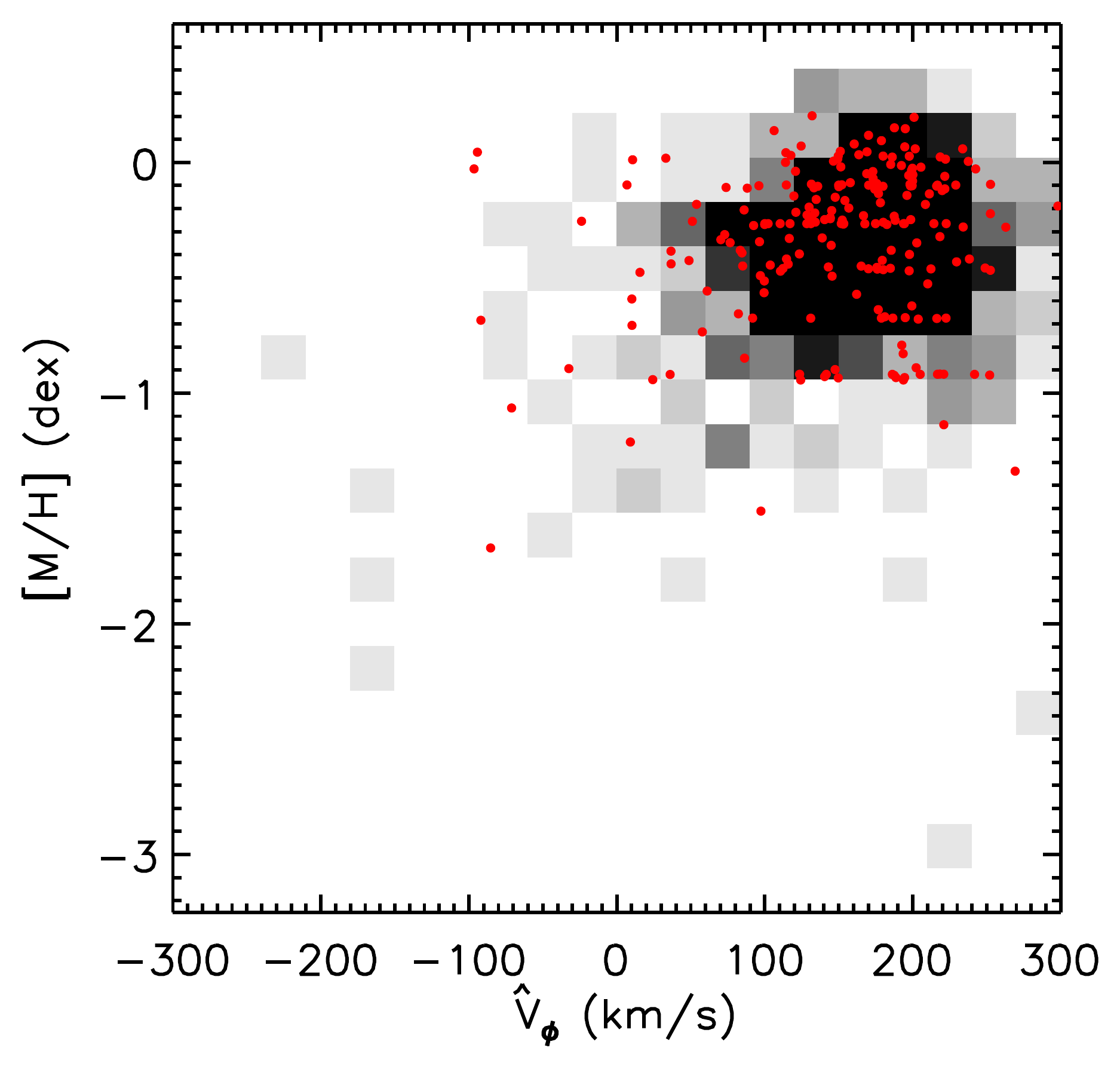} 
   \caption{ Upper panel:~Comparison between the observed  $\hat V_\phi$ distribution and that predicted by the Besan\c{c}on model for the Sextans line-of-sight. Lower panel: $\hat V_\phi$ versus metallicity for the observations (in red) and the Besan\c{c}on model. Very good agreement is found.}%
   \label{Fig:Vphi_sxt}
    \end{figure}

Finally, Fig.~\ref{Fig:Vphi_sxt} shows, for the Sextans line-of-sight, the comparisons between the model and the observations for the $\hat V_\phi$ distribution (upper panel) and for the relation between $\hat V_\phi$ and the metallicity (lower panel).  
It is clear that there is good agreement between the  observed data and the predictions of the Besan\c{c}on model. A KS-test indicates that both distributions are compatible, in the sense that the null hypothesis cannot be excluded with a probability of roughly $\sim$5\% (we note that within the Monte-Carlo realisations the probabilities of not rejecting the null hypothesis vary between 3\% and 15\%).
We remind the reader that $\hat V_\phi$ is not a good  estimator of the azimuthal velocity  for the Sculptor line-of-sight  and hence no comparison with the model can be made  in this direction. 

The general agreement between the observations and the model  -- which does not include any metallicity gradient for the thick disc, either radial or vertical --  leads to the conclusion that the decrease in metallicity with Z-height that is measured can be explained as reflecting not an intrinsic gradient, but rather a smooth transition from a thin-disc dominated population to a thick-disc dominated population  whose mean metallicity is around \meta$\sim -0.6$~dex and with a lag in orbital azimuthal velocity of $\sim 50$~\kms.  This result is in agreement with \cite{Cheng12} who detected no radial metallicity gradient for the thick disc stars, based on their analysis of spectra from the  SEGUE survey.  We note, however, that the error bars of our measurements are such that a gradient with amplitude as small as that found by \cite{Carrell12} ($0.02$~dex~kpc,$^{-1}$ based on SDSS photometric data for FGK dwarf stars) would not be detectable.    We conclude that a thick disc of constant mean metallicity and chemically symmetric relative to the Galactic plane, 
as is assumed in the Besan\c{c}on model, provides an adequate description of the data in these two lines-of-sight.
     
\subsection{The Carina line-of-sight: evidence of an accreted population or of the metal-weak thick disc?}

\begin{figure}[t]
   \centering
      \includegraphics[width=0.45\textwidth,angle=180]{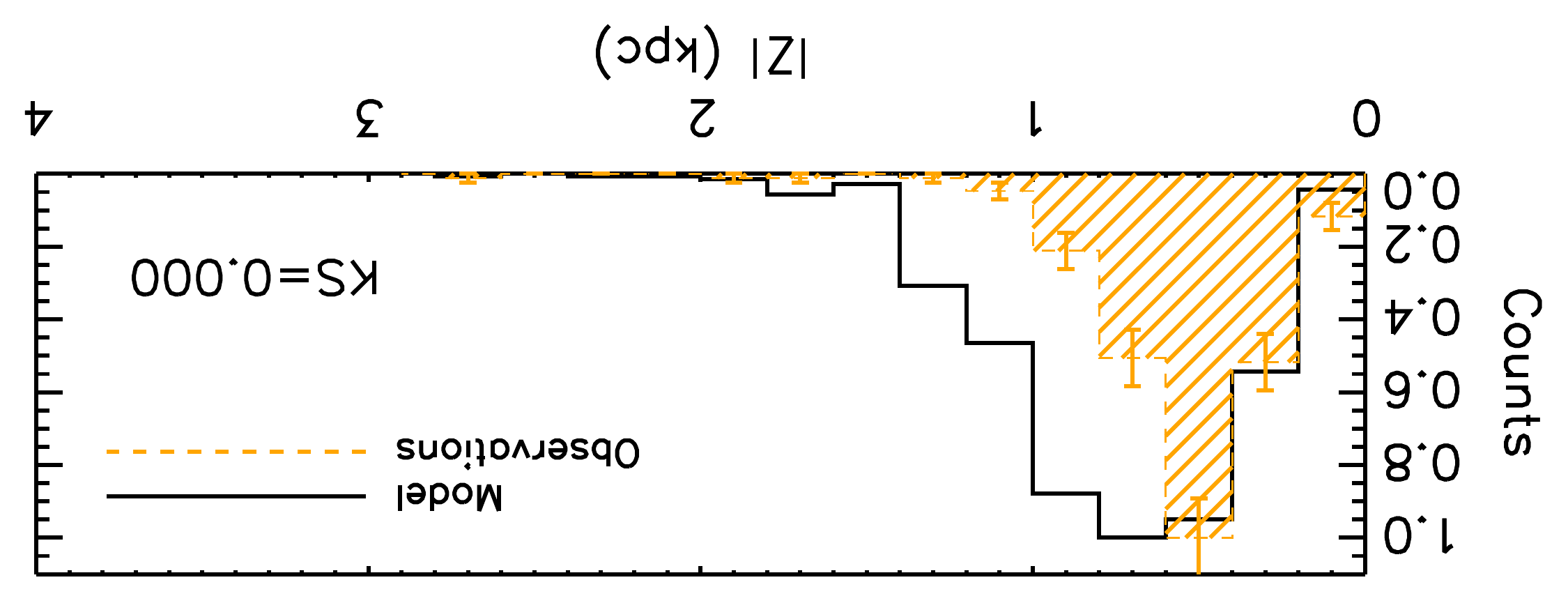} \\
\includegraphics[width=0.4\textwidth]{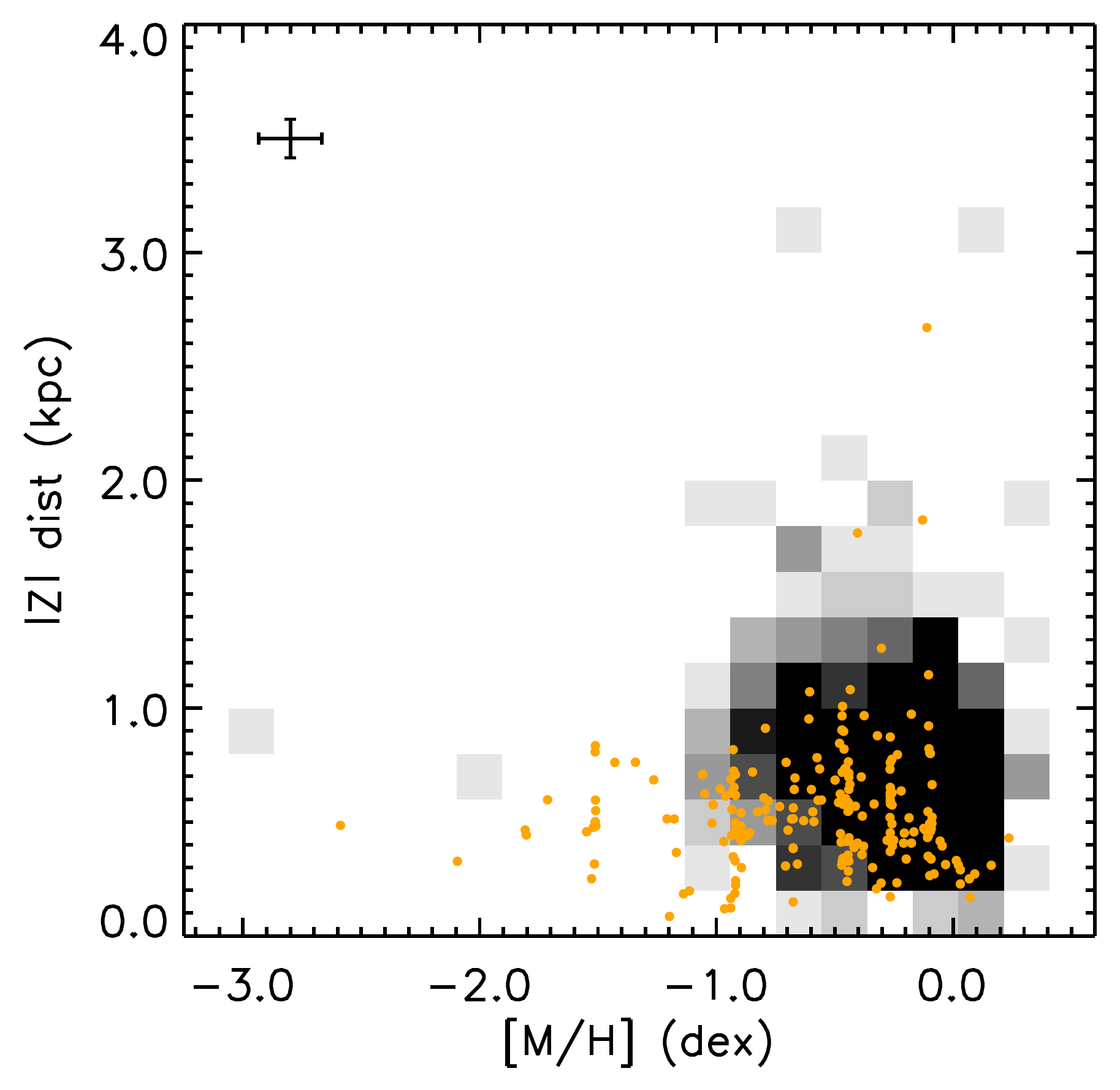} \\
\includegraphics[width=0.4\textwidth]{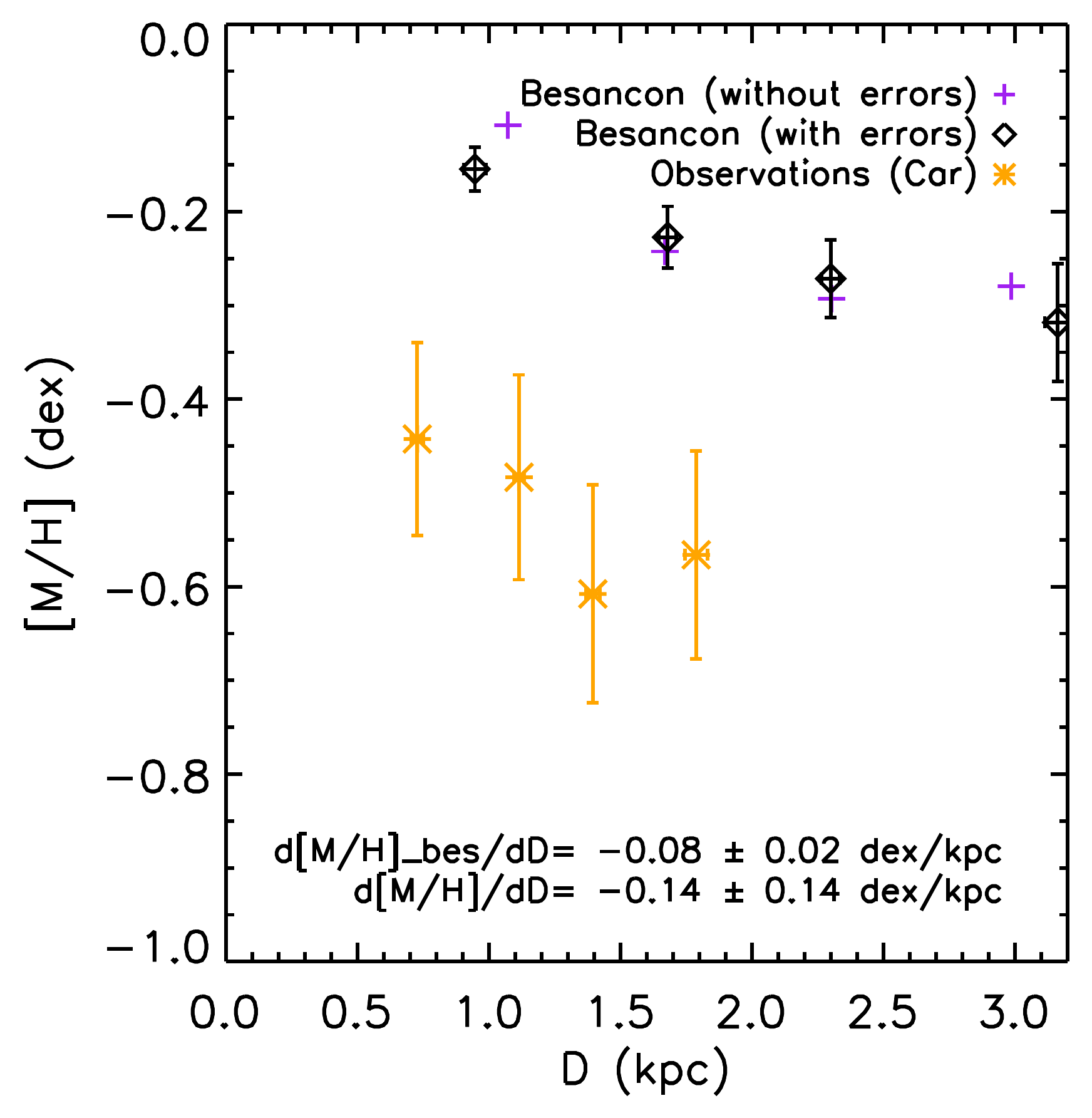} 
   \caption{ Upper panel:~Comparison of $|Z|-$distance distribution between the observations towards the Carina line-of-sight and the Besan\c{c}on model. Middle panel:~Metallicity versus distance above the Galactic plane for both the observed stars (orange filled circles) and simulated stars from the Besan\c{c}on model. Lower panel:~Derived metallicity gradients. The purple plus signs indicate the results obtained from the Besan\c{c}on model, assuming zero errors on position and metallicity.}
   \label{Fig:Comp_Car_Bes}
    \end{figure}

The roughly good agreement between the observations and the Besan\c{c}on model that is found for the lines-of-sight of Sculptor and Sextans, is not found in the case of the Carina line-of-sight, as it can be seen in Fig.~\ref{Fig:Comp_Car_Bes}. Indeed, for this lower Galactic latitude line-of-sight ($b = -22.2^\circ$), the relative population distributions show that the model lacks a significant number of stars with metallicities lower than [M/H]$<-1$~dex, at distances closer than 1~kpc from the plane. In addition, not only does the model under-predict the population of metal-poor stars, it over-predicts the population of  metal-rich stars at these distances. Further, the model over-predicts the number of stars of all metallicities at Z--distances greater than 1.5~kpc   (top panel of Fig.~\ref{Fig:Comp_Car_Bes}).

It is unlikely that the Galactic warp could  explain the difference between the model and the observations, since the data do not probe radial distance beyond 11~kpc, and the Galactic warp is expected to have an influence only at distances greater than 12~kpc \citep[e.g.][]{Momany06}. The mis-match between model and data can therefore be  due to one of more of the following:  the assumption of an incorrect value for the extinction; the presence of metallicity gradients in the thin disc (both radially and vertically), neglected in the model; the adoption in the model of an incorrect value for the relative normalisation of thick and thin discs; or, lastly, the presence of an {\it extra} population or structure, not included in the model, such as due to accretion or a moving group.

       \begin{figure}[t]
   \centering
\includegraphics[width=0.45\textwidth,angle=180]{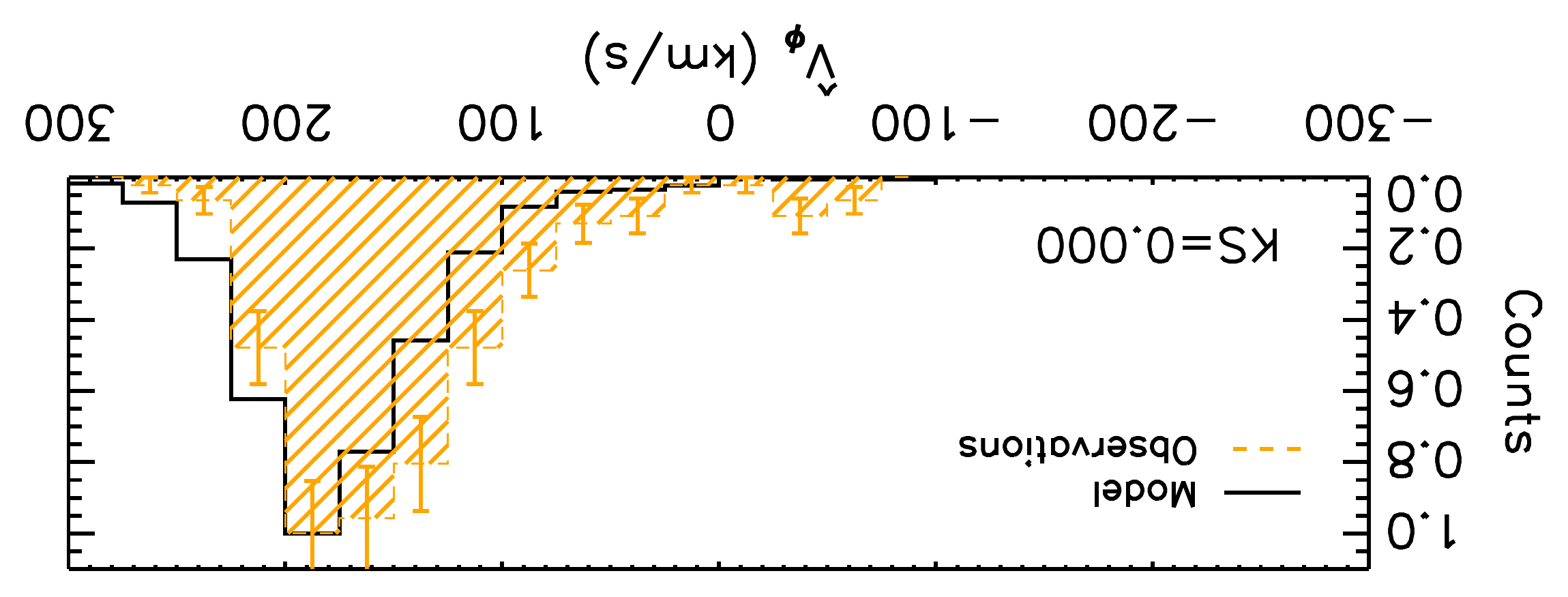} \\
\includegraphics[width=0.45\textwidth]{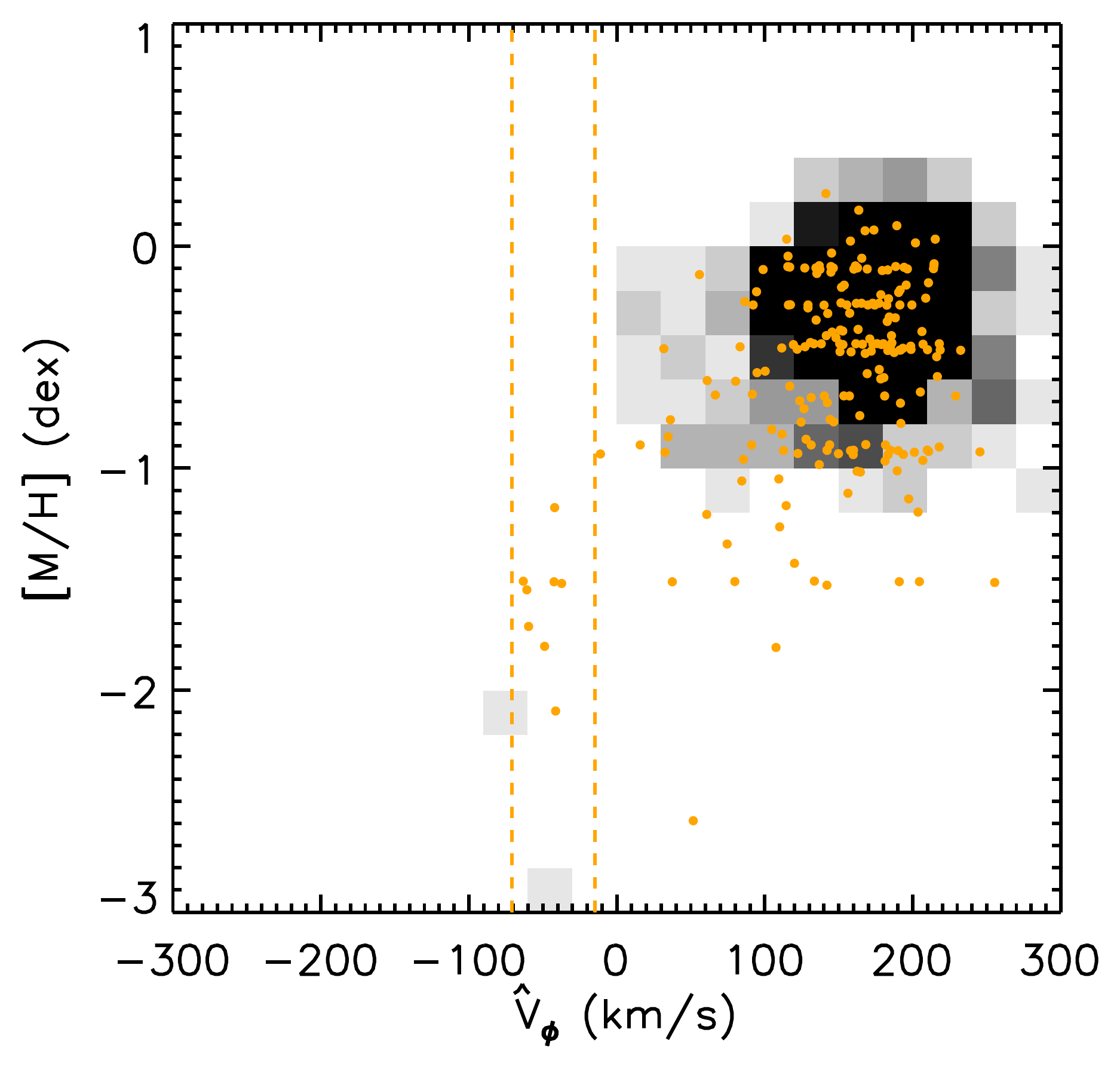} 
   \caption{Upper panel:~Comparison between the derived $\hat V_\phi$ distribution of the observed sample with that  predicted by the Besan\c{c}on model for the Carina line-of-sight. Lower panel:  $\hat V_\phi$ versus metallicity for the observations (filled orange circles) and for the Besan\c{c}on model. The vertical dashed lines indicate the expected range for true members of the Carina dSph.}%
   \label{Fig:Vphi_car}
    \end{figure}

Figure~\ref{Fig:Vphi_car} is similar to Fig.~\ref{Fig:Vphi_sxt} but for the line-of-sight of Carina. It is apparent  that the difference in the azimuthal velocity distribution is not simply  a zero-point offset, but, as seen on the lower panel of Fig.~\ref{Fig:Vphi_car}, reflects the absence in the models of  the low metallicity population that is observed relatively close to the Galactic plane. This conclusion is supported by K-S tests on the line-of-sight and azimuthal velocities distributions which exclude the null hypothesis that the simulations come from the same underlying distribution as do  the observations. The vertical dashed lines in the lower panel of Fig.~\ref{Fig:Vphi_car}  correspond to the expected range for stars of the background Carina dSph; contamination of the sample by background stars should not be important. 

\begin{figure}[t]
   \centering
\includegraphics[width=0.4\textwidth]{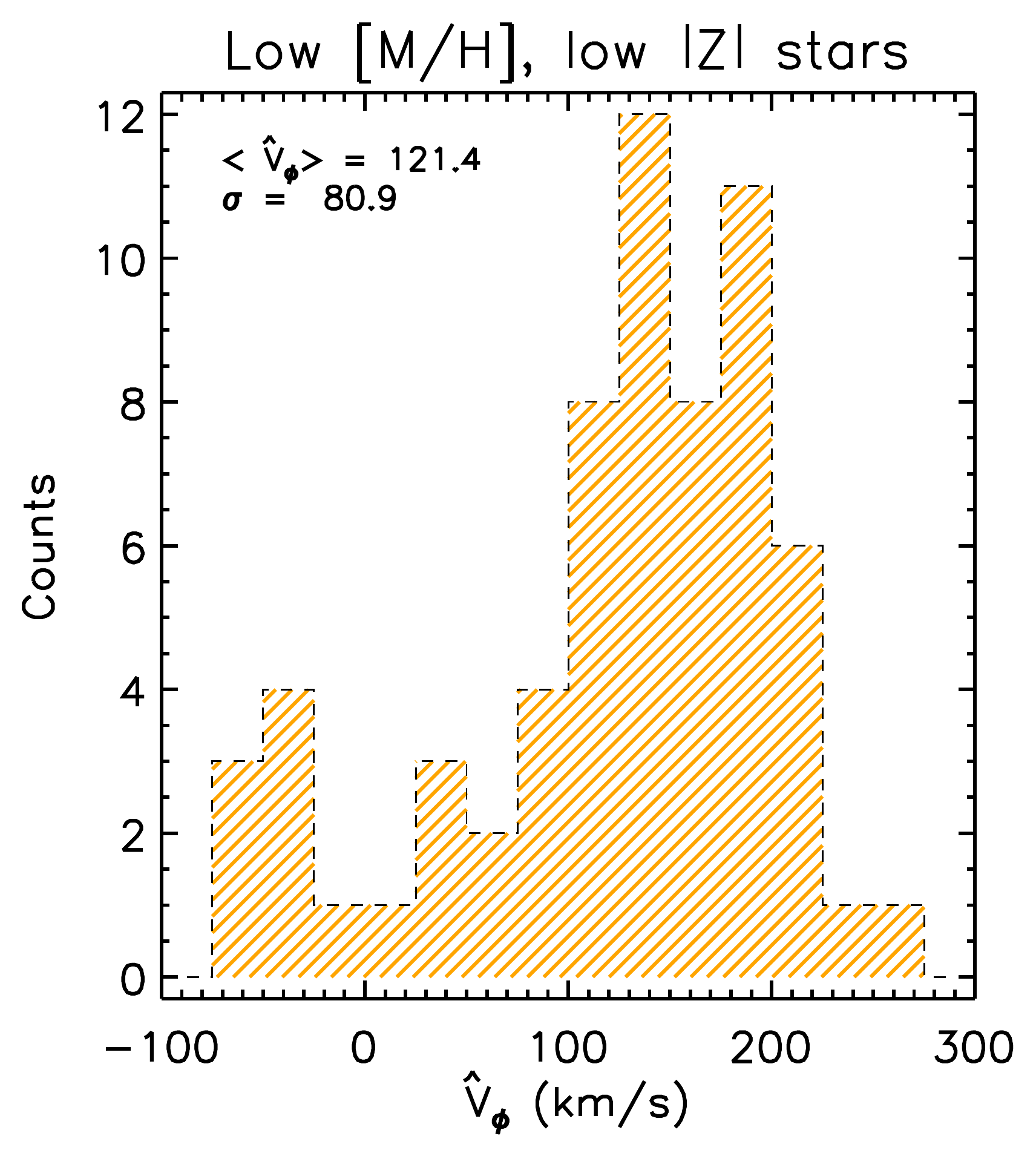} 
   \caption{Azimuthal velocity estimator for the stars towards the Carina line-of-sight that have both low metallicity and are located close to the Galactic plane (not predicted by the Besan\c{c}on model). The histogram has a peak at $\hat V_\phi \approx V_\phi \sim 120$~\kms, an intermediate value between  those expected for the canonical thick disc and the halo.  }%
   \label{Fig:Car_lowVphi}
    \end{figure}

This low-metallicity population appears to be confined close to the plane.  We investigated it further by selecting all stars for which we derived both a metallicity lower than $-0.9$~dex and a $Z-$distance smaller than 0.75~kpc. 
The distribution of the projected azimuthal velocity ($\hat V_\phi$) for these stars is shown in Fig.~\ref{Fig:Car_lowVphi} (as shown in the bottom panel of Fig.~\ref{Fig:Vphi_estim}, this a good estimator of the true orbital azimuthal velocity). The distribution peaks at $\sim120$~\kms, well below the value of the thin disc ($\hat V_\phi \approx V_\phi \sim 210$~\kms), and also below that of the canonical thick disc ($V_\phi \sim170$~\kms)  but above the mean expected for the stellar halo ($V_\phi \sim 0$~\kms).  The measured velocity dispersion (80~\kms) is also intermediate between the expected value of the halo ($\sim 100$~\kms) and that of the canonical thick disc ($\sim 50$~\kms).   These kinematic properties of this population are in good agreement with the findings of \cite{Wyse06}, who analysed the same dataset  and identified an 'extra' component. We note however that unlike \cite{Wyse06}, in the present analysis we have derived precise distance estimations  ($|Z|<1$~kpc) and metallicity measurements ([M/H]$<-1$~dex) and thus can make a more robust characterisation of this population.

\begin{figure}[t]
   \centering
   \includegraphics[width=0.4\textwidth]{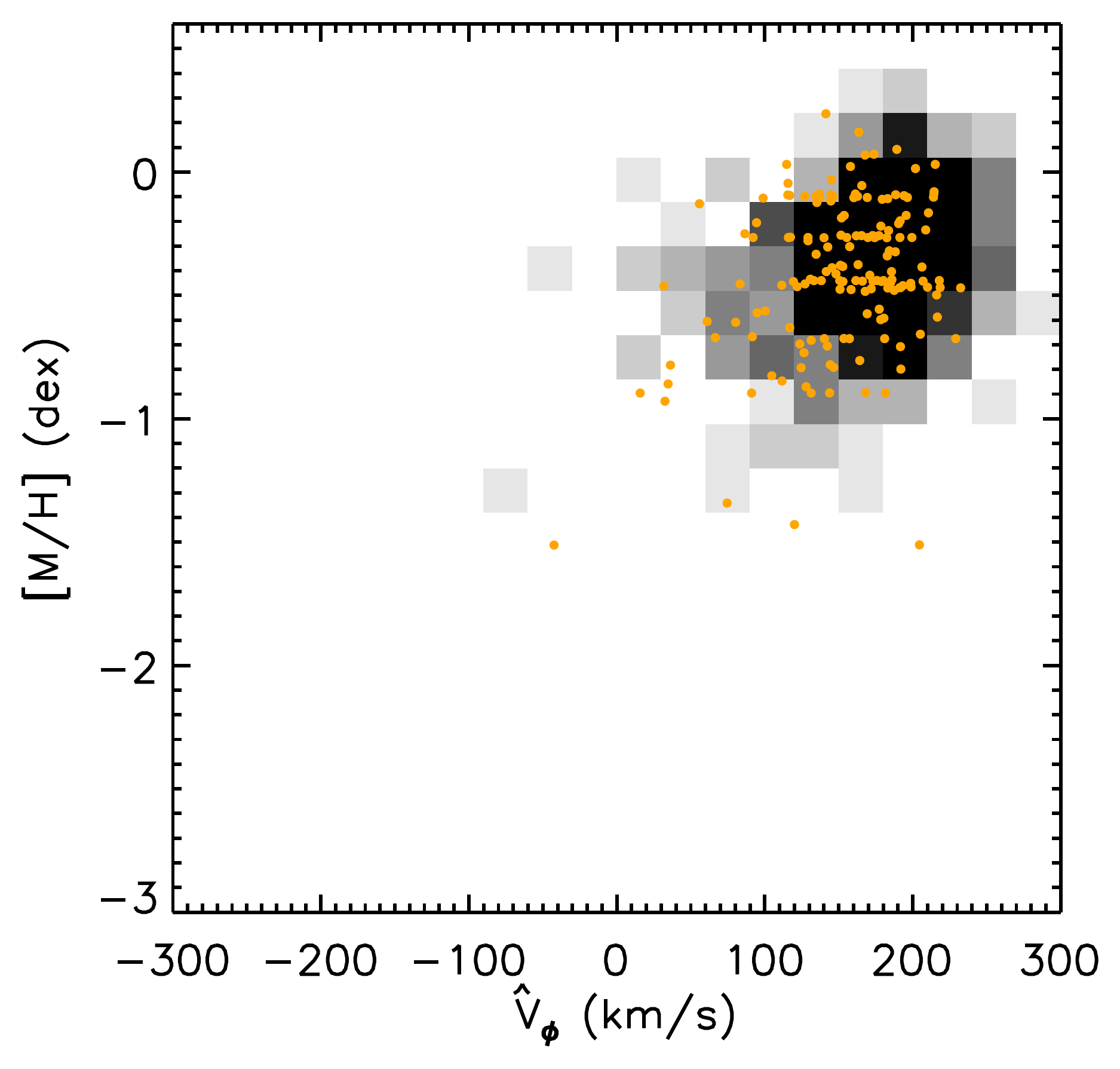} \\
\includegraphics[width=0.4\textwidth]{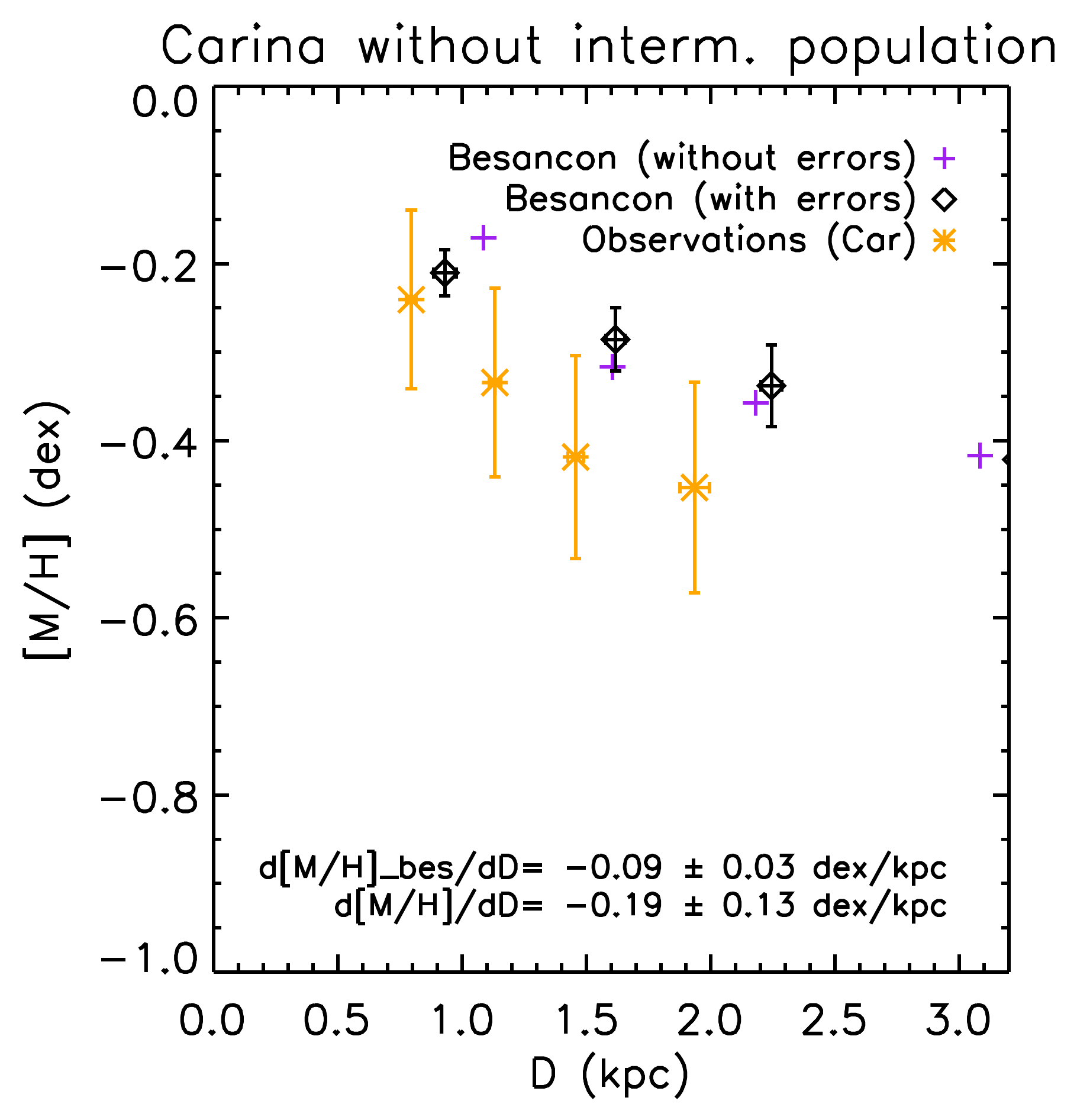} 
   \caption{Observed and predicted gradients for the Carina line-of-sight, without the low-altitude, low metallicity stars  of Fig.~\ref{Fig:Car_lowVphi} ($\meta < -0.9$~dex \&  $|Z| < 0.75$~kpc).    The purple plus signs indicate the results obtained from a Besan\c{c}on model without any errors on the positions nor the metallicity.  }%
   \label{Fig:Comp_Car_Bes2}
    \end{figure}

The kinematic characteristics of these stars, combined with the low value of their mean metallicity ($\sim -1.5$~dex, typical of stars in the stellar halo or in dSph) indicate that this population does not belong to the canonical thick disc.  However, their locations, close to the Galactic plane, argue against membership of these stars in the stellar halo. Further, both the position and the metallicity of this {\it extra} population rule out the possibilities that the disagreement between the model and the observations is caused purely by either the adoption of an erroneous value for the interstellar extinction in our analysis or by different radial and/or vertical metallicity gradients than those assumed in the Besan\c{c}on model \citep[see, for example,][for values of radial metallicity gradients in the thin disc ranging from $-0.046$~dex~kpc$^{-1}$ to $-0.085$~dex~kpc$^{-1}$]{Rudolph06,Pedicelli09,Gazzano13}.
Indeed, we found good agreement between the model predictions and our observational data, other than the presence of this set of low angular momentum stars only in the observations.  This is demonstrated in Fig.~\ref{Fig:Comp_Car_Bes2}, which shows the analogues of the lower panels of Fig.~\ref{Fig:Comp_Car_Bes} and Fig.~\ref{Fig:Vphi_car}, but now with the members of the {\it extra\/} population removed from the observed dataset. We proceeded by first excluding these stars from the dataset and recomputing an appropriate simulated catalogue from the Besan\c{c}on model. We then derived the metallicity gradients, given $5\times 10^3$~Monte-Carlo realisations.
As seen in Fig.~\ref{Fig:Comp_Car_Bes2},  the agreement is much improved, especially for the closest stars, with  the $\hat V_\phi$ versus \meta\ distribution similar to that predicted by the model\footnote{However, a K-S test on the velocity distributions still rejects the null hypothesis, but the probability that they are drawn  from the same underlying population has increased.}.  In addition, the new normalisation of the model in the different magnitude bins removes the over-prediction of stars far from the plane (see upper panels of Fig.~\ref{Fig:Comp_Car_Bes}).
We note that even after pruning the sample, the derived metallicity gradient from the observations is still steeper than the predicted one.

\section{Discussion: accreted population, metal-weak thick disc or moving group?}
\label{Sect:Discussion}

Moving groups, localised in phase space and metallicity have their origin from disrupted star clusters, satellite accretion or due to
resonances between disc stars and the Galactic bar and/or transient spiral arms \citep{Antoja10}. In the case of disrupted stellar clusters, the stars composing the moving group are usually young, with typical ages of less than 1~Gyr. It hence seems unlikely, given the low metallicity
and the relatively wide (1~dex) metallicity range of the stars we have identified in this population along the  Carina line-of-sight, that this origin could be viable.

On the other hand, a moving group with old ages (in order to be compatible with metallicities lower than --1~dex) can be produced from
the resonances between the spiral arms and the Galactic bar. 
Nevertheless, it is unlikely that a moving group with such an high azimuthal velocity dispersion would have formed. Indeed, in order for a group of stars to migrate outwards, similar velocities are required, in order to be simultaneously in resonance with the Galactic
bar and the spiral arms.

An alternative explanation was proposed by \cite{Gilmore02} and \cite{Wyse06}. These authors have argued that in the general framework of   thick-disc formation through heating of a pre-existing thin disc by accretion of a massive satellite galaxy  at a redshift  $z \sim 2$,  stellar debris from the satellite would likely be of low metallicity and on  intermediate angular momentum orbits.   In that case, the fact that these stars are not observed towards the other line-of-sight of Sextans and Sculptor could be due to the initial orbit of the satellite and peri-centre passage at which the observed stars were removed from the satellite. Nevertheless, this argument seems to be refuted by the fact that \cite{Kordopatis11b}, who observed towards $(l,b)=(277^\circ,47^\circ)$, did not find any evidence of such a population, despite the fact that distances up to 5--6~kpc above the plane were covered. In addition, the azimuthal velocity dispersion of the {\it extra\/} population seems too  high for  a localised stellar stream, further challenging  this scenario.
  We note, however, that  \cite{Mizutani03} have suggested that a putative  dwarf spheroidal progenitor of $\omega-$Cen could have deposited debris with similar low angular momentum and metallicity. \\

Finally, this population could be related to the Metal-Weak Thick Disc (MWTD, hereafter). The existence of such a metal-poor tail for the thick disc has already been discussed in \cite{Morrison90, Chiba00,Carollo10, Ruchti11} and represents an additional component associated with the canonical thick disc, extending to metallicities as low as --2~dex.  The {\it extra\/} population identified  in this study has a mean metallicity of  $\sim -1.4$~dex and mean $\hat V_\phi \approx V_\phi \sim 120-130$~\kms. Assuming that the canonical thick disc has a mean
metallicity of $-0.5$~dex and a mean $V_\phi \sim 170$~\kms, this would imply that the MWTD component that we observed follows  the proposed \citet{Spagna10, Kordopatis11b, Lee11} orbital rotation-metallicity correlation of $\partial V_\phi
/ \partial \meta \approx 40-50$~\kms~dex$^{-1}$.  Following this line of argument, the {\it extra\/} population is simply the metal-poor tail of the canonical thick disc.

The question then arises as to why we do not observe this MWTD population towards the other Galactic directions. One  obvious reason could be  possible substructure within the thick disc. Moreover, the Carina line-of-sight is at a lower Galactic latitude and thus has a longer path length through the thick disc which should lead to a larger number of foreground stars, consequently detecting more stars belonging to the metal-weak tail of the thick disc.  Indeed, assuming no variation in the scale-height of the thick disc for the probed Galactocentric radii, the Carina line-of-sight should pass through  $\sim 1.8$ times more thick disc stars than the Sextans line-of-sight, and $\sim 2.6$ times more thick disc stars than the Sculptor line-of-sight (as determined by the ratio of the {\it sine} of their Galactic latitudes). It is hence possible that the MWTD would not have been detected in the other line-of-sight due to small  number statistics.

In order to further investigate this possibility, we first assume that the contribution of the stellar halo can be neglected in both  the lines-of-sight of Carina and Sextans, and compute the ratio of the number of candidate MWTD stars (\meta$<-1.05$~dex, $|Z|<2$~kpc) to the total number of thick disc stars (\meta$<-0.5$~dex, $|Z|<2$~kpc).  This gives roughly 28\% (26 stars out of  92) and 13\% (7 stars out of  52) of the targets belong to the MWTD, towards the Carina and Sextans line-of-sight, respectively. These numbers are in good agreement, after multiplying the result for the Sextans line-of-sight by the corrector factor of 1.78 due to the different latitudes.  It hence seems plausible to argue that the Sextans line-of-sight has a similar proportion of MWTD stars, which have barely been observed due to the low normalisation  of the metal-weak tail of the thick disc.

The contribution of the stellar halo cannot be neglected for the Sculptor line-of-sight since in this direction the foreground stars are at larger  distances from the plane compared to the other lines-of-sight.  Indeed, a similar star-count ratio as in the other lines-of-sight, based on just the metallicities and assuming entirely thick disc, would indicate that roughly 40\% (22 stars out of 55) should belong to the MWTD.
Unfortunately, without the azimuthal velocity,  halo stars are difficult to disentangle from the rest of the Galactic components. However, in this line-of-sight the radial velocity is a good proxy for the vertical velocity, $V_Z$, and this may be used as a basis for separation of the thick disc from the halo. Adopting the criterion that  thick disc stars have $|V_{\rm rad}|<50$~\kms), the  MWTD fraction drops to $\sim 25$\%, indicating that indeed  a considerable number  of halo stars contaminates the sample towards this line-of-sight.

   
   \section{Conclusions}
   \label{Sect:conclusions}

We have taken advantage of spectroscopic surveys that aimed to study member stars in four dwarf spheroidal galaxies (Fornax, Sculptor, Sextans and Carina) to instead study  stars of the Milky Way Galaxy, observed serendipitously as foreground 'contaminants' in the sample. We selected the last three directions and extracted  the foreground stars in order to  investigate the chemical properties of our Galaxy towards
 the outer discs.

We ran the medium-resolution spectra through an automated pipeline in order to derive estimated values for the stellar atmospheric parameters and then we projected the output on theoretical isochrones, to infer the stellar line-of-sight distances. We compared the observational results with the predictions of the Besan\c{c}on Galactic model \citep{Robin03} and showed that towards the directions of
Sculptor and Sextans, the decreases in mean metallicity with distance that were found are consistent with zero intrinsic vertical and radial
metallicity gradients in the thick disc, reflecting rather the change in population mix (thin disc -- thick disc) with distance.

      Furthermore, we confirmed the existence of an unpredicted {\it extra} population towards the line-of-sight of Carina, previously proposed by \citet{Wyse06}. This population is observed close to the plane ($|Z|<1$~kpc), has a mean metallicity around $\sim -1.4$~dex, and an intermediate mean azimuthal velocity ($V_\phi \sim 120$~\kms). It follows the well established correlation between the metallicity and the azimuthal orbital velocity of thick-disc stars, and hence is likely to be associated with the metal-weak thick disc.  We note, however, that future tests will need to involve 3D kinematics (incorporating proper motions) in order to derive the velocity ellipsoid of these stars, and estimate their scale-length and scale-height. This will allow a more comprehensive comparison between this {\it extra\/} population and  the canonical thick disc, and hence  establish more robustly if it  is indeed the metal-poor tail of the canonical thick disc or a different component.

Finally, we showed that if we remove this {\it extra} population from our analysis of the data towards Carina, then the Besan\c{c}on model -- which assumes no gradients in the metallicity distribution of the thick disc -- agrees, within the errors, with the observations, as it did towards the two other studied lines-of-sight. This result implies  that the thick disc has approximately constant mean metallicity, at least within several kpc of the Sun, above and below the Galactic Plane.

\bibliographystyle{aa} 
\bibliography{Dart} 

\begin{acknowledgements} 
J.~E.~Norris is warmly thanked for his careful reading of the manuscript and his useful comments. The anonymous referee is also thanked for helping us improve the the quality of our paper. RFGW acknowledges support from National Science Foundation grants
AST-0908326 and CDI-1124403. E.S. gratefully acknowledges the Canadian Institute for Advanced Research (CIfAR) Global Scholar Academy and the Canadian Institute for Theoretical Astrophysics (CITA) National Fellowship for support.
G. B. gratefully acknowledges support through a Marie-Curie action Intra European Fellowship, funded from the European Union Seventh Framework Program (FP7/2007-2013) under grant agreement number PIEFGA-2010-274151.

 \end{acknowledgements}

\end{document}